\documentclass[aps,pra,twocolumn,amsmath,amssymb,floatfix,reprint,footinbib,superscriptaddress,longbibliography,showkeys]{revtex4-1}
\usepackage{dcolumn}
\usepackage{graphicx}
\usepackage{mathrsfs}
\usepackage{mdwlist}
\usepackage{subfigure}
\usepackage{booktabs}
\usepackage{multirow}
\usepackage{amsmath}
\usepackage{textcomp}
\usepackage{upgreek}
\usepackage{dsfont}
\usepackage{amstext}
\usepackage{amssymb}
\usepackage{amsbsy}
\usepackage{appendix}
\usepackage{soul}
\usepackage{threeparttable}
\usepackage{bbm}
\usepackage{bm}
\usepackage{amsthm}
\usepackage{graphicx}
\usepackage{textcomp}
\usepackage{xcolor}
\usepackage{soul}
\usepackage{multirow}
\usepackage{color}
\usepackage{diagbox}
\usepackage{braket}
\usepackage[colorlinks,citecolor=blue]{hyperref}
\definecolor{Dgreen}{RGB}{0, 100, 0}
\usepackage{url}
\usepackage[colorlinks]{hyperref}
\hypersetup{%
	plainpages=true,
	breaklinks=true, 
	hypertexnames=false, 
	pageanchor=true,
	colorlinks=true,
	linkcolor={blue},
	citecolor={blue},
	urlcolor={blue},
	anchorcolor={black}
}

\begin{document}

\title{Noise-resilient nonadiabatic geometric quantum computation for bosonic binomial codes}

\author{Dong-Sheng Li}
\affiliation{Fujian Key Laboratory of Quantum Information and Quantum Optics, College of Physics and Information Engineering, Fuzhou University, Fuzhou, Fujian 350108, China}

\author{Yang Xiao}
\affiliation{Fujian Key Laboratory of Quantum Information and Quantum Optics, College of Physics and Information Engineering, Fuzhou University, Fuzhou, Fujian 350108, China}

\author{Yu Wang}
\affiliation{School of Physics, Hangzhou Normal University, Hangzhou 311121, China}%

\author{Yang Liu}
\affiliation{Center for Joint Quantum Studies and Department of Physics, School of Science, Tianjin University, Tianjin 300350, China}%

\author{\\Zhi-Cheng Shi}
\affiliation{Fujian Key Laboratory of Quantum Information and Quantum Optics, College of Physics and Information Engineering, Fuzhou University, Fuzhou, Fujian 350108, China}

\author{Ye-Hong Chen}\thanks{yehong.chen@fzu.edu.cn}
\affiliation{Fujian Key Laboratory of Quantum Information and Quantum Optics, College of Physics and Information Engineering, Fuzhou University, Fuzhou, Fujian 350108, China}
\affiliation{Institute of Quantum Science and Technology, Yanbian University, Yanji 133002, China}
\affiliation{Quantum Information Physics Theory Research Team, Center for Quantum Computing, RIKEN, Wako-shi, Saitama 351-0198, Japan}%

\author{Yi-Hao Kang}\thanks{yihaokang@hznu.edu.cn}
\affiliation{School of Physics, Hangzhou Normal University, Hangzhou 311121, China}%

\author{Yan Xia}\thanks{xia-208@163.com}
\affiliation{Fujian Key Laboratory of Quantum Information and Quantum Optics, College of Physics and Information Engineering, Fuzhou University, Fuzhou, Fujian 350108, China}

\affiliation{Institute of Quantum Science and Technology, Yanbian University, Yanji 133002, China}

\begin{abstract}

The binomial code is renowned for its parity-mediated loss immunity and loss-error recoverability, while geometric phases are widely recognized for their intrinsic resilience against noise. Capitalizing on their complementary merits, we propose a noise-resilient protocol to realize Nonadiabatic geometric quantum computation with binomial codes in a superconducting system composed of a microwave cavity 
dispersively coupled to a 
qutrit. The control field 
 is designed by 
integrating reverse engineering and optimal control. This design provides a customized control protocol featuring strong error-tolerance and inherent noise-resilience.
Using experimentally accessible parameters in superconducting systems, numerical simulations show that the protocol yields relatively high average fidelity for geometric quantum gates based on binomial code, even in the presence of parameter fluctuations and decoherence. Thus, this protocol may provide a practical approach for realizing reliable Nonadiabatic geometric quantum computation with binomial codes in current technology.


\end{abstract}

\maketitle

\section{Introduction}\label{I}

Quantum computation \cite{RevModPhys.68.733,PhysRevLett.79.325,Jones2000,PhysRevA.93.012306,PhysRevA.95.022319} leverages the superposition and entanglement of qubits to deliver significant computational advantages over its classical counterpart for intractable tasks, including prime factorization \cite{S0036144598347011,Vandersypen2001,PhysRevLett.108.130501,PhysRevA.105.062455} and large-scale searching \cite{PhysRevA.62.062303,WOS001177135800178}. However, the construction of fully functional quantum computers still faces significant challenges due to local parameter fluctuations, systematic errors, and decoherence in experiments \cite{Kang_2023,PhysRevA.96.022309,PhysRevA.97.042336,PhysRevA.100.043413}. Thus, suppressing and correcting errors caused by these defects have become a critical issue that urgently needs to be addressed in the field of quantum computations. \cite{PhysRevA.92.052302,PhysRevApplied.10.054051,PhysRevA.96.032341,PhysRevLett.123.200502}

To address the aforementioned challenges, various approaches have been proposed \cite{PhysRevA.52.R2493,PhysRevA.86.032324,PhysRevLett.119.180509,Kandala2019}, such as variational quantum algorithms \cite{Peruzzo2014,Kandala2017}, optimal control \cite{PhysRevLett.125.250403,doi:10.1126/sciadv.adr0875}, etc. Among these paradigms for mitigating experimental imperfections, Nonadiabatic geometric quantum computation (NGQC) exhibits distinct advantages \cite{Sjqvist_2012,PhysRevLett.109.170501,PhysRevA.92.022320}. On the one hand, NGQC capitalizes on the global attributes of geometric phases \cite{10.1098/rspa.1984.0023,PhysRevLett.58.1593,PhysRevA.95.032311,PhysRevA.101.022330}, providing an effective pathway to improve the robustness of quantum gates. On the other hand, in contrast to traditional adiabatic methods, it overcomes the strict time constraints imposed by adiabatic conditions, significantly reducing operational duration while effectively suppressing 
the influence of decoherence \cite{PhysRevA.95.062310,PhysRevA.98.052315}. This breakthrough partially resolves the core challenge that quantum states are highly susceptible to environmental interference. Moreover, NGQC demonstrates exceptional compatibility, enabling deep integration with reverse engineering \cite{PhysRevA.101.032322}, optimal control \cite{PhysRevA.100.012329,PhysRevLett.123.100501}, dynamical decoupling \cite{PhysRevA.103.012205}, and other cutting-edge methodologies \cite{Zhou15,PhysRevA.97.022335}. Consequently, the robustness of NGQC against parameter fluctuation and decoherence can be further enhanced.

Beyond the dynamic control strategies represented by NGQC, fault-tolerant quantum codes \cite{PhysRevX.10.011058,PhysRevA.97.032346,Heeres2017,CAI202150,vf7v-cpq9,PhysRevResearch.4.013233} constitute another cornerstone for reliable quantum computation, where bosonic codes play an irreplaceable role \cite{PhysRevX.10.011058,PhysRevA.97.032346,PhysRevResearch.4.013233}. The bosonic codes provide a highly promising solution for constructing stable logical qubits. Unlike physical codes, which rely on the entanglement of multiple physical qubits for quantum error correction, bosonic codes use the infinite-dimensional Hilbert space of a bosonic system to encode logical information \cite{PhysRevA.56.1114,PhysRevA.64.012310,PhysRevLett.116.140502,PhysRevX.9.041053,PhysRevLett.126.023602}. The unique encoding architecture of bosonic codes ensures the number of error channels invariant with increasing encoding redundancy \cite{doi:10.1126/science.1243289,Albert_2019}, 
cutting the overheads of quantum hardware fabrication and manipulation. Meanwhile, the error channel of bosonic codes is determined solely by the interaction between a single bosonic mode and the environment \cite{PhysRevLett.119.030502,CAI202150,PhysRevX.6.031006,PhysRevLett.124.120501}.
This advantage facilitates efficient, straightforward error detection and correction. 

The binomial code \cite{PhysRevX.6.031006}, is an important type of bosonic codes, where quantum information is encoded on the logical states
\begin{eqnarray}\label{e3}
|\mathbbmss{O}\rangle=\frac{1}{\sqrt{2}}\left(|0\rangle+|4\rangle\right),\   |\mathbbm{1}\rangle=|2\rangle,
\end{eqnarray}
with the Fock state $|n\rangle$ ($n=0,1,2,\dots$) of a bosonic mode. 
In such encoding, a photon loss error will map the logical states to an auxiliary subspace with odd parity (odd-number photon), which is 
orthogonal to the code subspace with even parity (even-number photon). Since the bosonic code satisfies the Knill-Laflamme condition $\langle\mathbbm{\varrho}|a^{\dag}a|\mathbbm{\varrho'}\rangle=2$ ($\varrho, \varrho'=\mathbbmss{O}, \mathbbm{1}$), the quantum information 
will not be deformed when a photon loss happens \cite{PhysRevA.54.3824,PhysRevA.55.900}. Moreover, the photon loss error can also be detected via a photon number parity measurement and restored through unitary operations \cite{PhysRevX.6.031006}.
The advantages of binomial codes have driven ‌sustained interest in their application to quantum computation, both theoretically and experimentally \cite{Chou2018,Hu2019,Gao2019,PhysRevLett.124.120501,Ma2020}. These remarkable works have further inspired researchers to integrate binomial code into NGQC  \cite{PhysRevResearch.3.033275,Kang22}. However, 
they still face some challenges in experimental realization, stemming from the requirements of ultra-strong coupling \cite{PhysRevResearch.3.033275} and nonlinear interactions \cite{Kang22}.




In this paper, we propose a more experimentally friendly and noise-resilient protocol for realizing NGQC based on binomial codes. The physical model consists of a microwave cavity encoding the binomial code qubit and a superconducting qutrit serving as an auxiliary. The transition between the highest and intermediate levels of the qutrit is dispersively coupled to the microwave cavity and simultaneously driven by a detuned external microwave field. This yields an effective Hamiltonian wherein the cavity component can be diagonalized in a displaced Fock state basis when the qutrit occupies the intermediate level. The transition between the lowest and intermediate levels is driven by a microwave field with three frequency components, establishing a channel between the displaced vacuum state and each Fock state employed in the binomial code via frequency matching. This channel enables logical state manipulation without requiring complex nonlinear interactions or ultra-strong coupling.

Control field is designed through the integration of reverse engineering and optimal control theory. The reverse engineering begins by identifying a Hermitian dynamical invariant operator associated with the effective Hamiltonian \cite{PhysRevA.101.012345,PhysRevA.109.022437,wfj8-tgjz}. The eigenstates of this dynamical invariant define the evolution paths for geometric gate operations \cite{PhysRevA.97.012333,PhysRevLett.126.023602,202200348,Yu25}. Subsequently, optimal control theory guides the design of control parameters. By minimizing a cost function that quantifies systematic error sensitivity, we derive an error-tolerant control protocol for implementing nonadiabatic geometric quantum gates with binomial codes.


At last, the impacts of random noise and decoherence are also studied via numerical simulations. The results demonstrate that the protocol can produce an acceptable average fidelity for geometric quantum gates in the presence of these disturbing factors. Therefore, the protocol provides useful perspectives for the realization of robust nonadiabatic geometric quantum compuation with binomial codes.

The paper is organized as follows. In Sec.~\ref{II}, we 
introduce the physical model and derive the effective Hamiltonian. In Sec.~\ref{III}, we elaborate the construction of evolution paths and the design of control parameters for binomial-code NGQC assisted by reverse engineering and optimal control theory. In Sec.~\ref{IV}, we perform numerical simulations to validate the feasibility of the protocol and its robustness against systematic errors in control fields. We also estimate the performance of the protocol in the presence of random noise and decoherence (including the energy relaxation, dephasing of the qutrit, and the decay of the cavities). Finally, the conclusion is given in Sec.~\ref{V}.

\section{PHYSICAL MODEL AND EFFECTIVE HAMILTONIAN} \label{II}

\begin{figure}
  \centering
  \includegraphics[width=8cm]{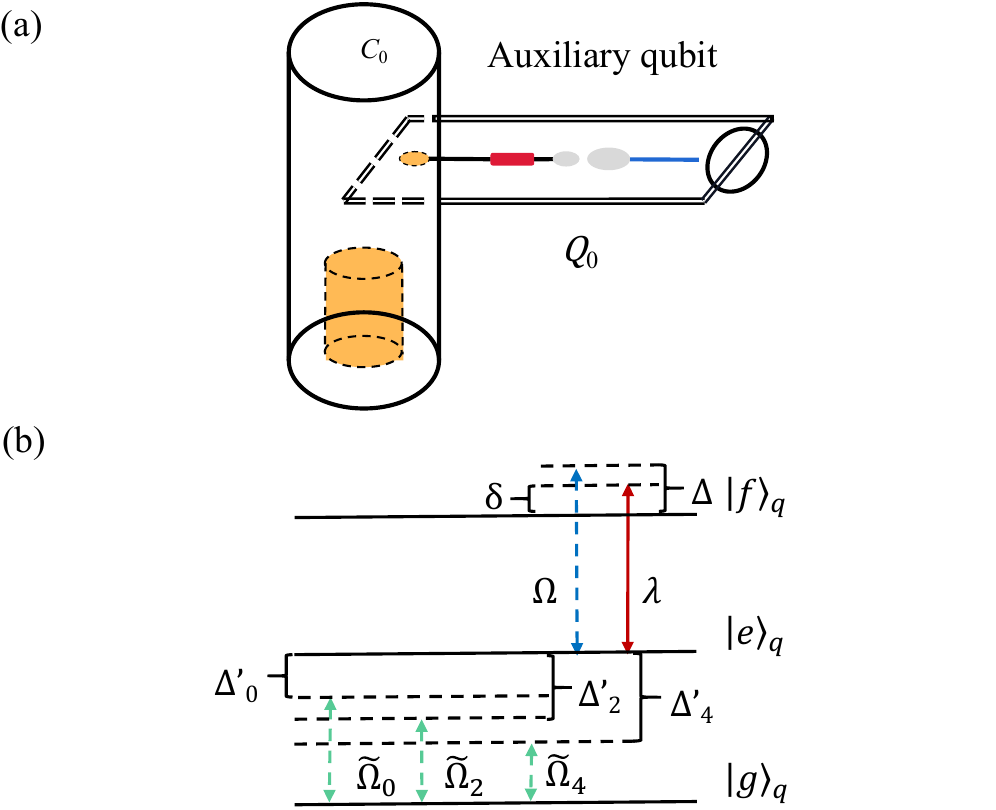}\caption{(a) Configuration of the superconducting system, which consists of a 3D superconducting microwave cavity $C_0$ capacitively coupled to a superconducting flux qutrit (a three-level system) $Q_0$, where the orange cylinder represents the coaxial cavity of the 3D superconducting microwave cavity, the yellow ellipse indicates the coupling port between the cavity and the right-side superconducting qubit, and the red rectangle denotes the Josephson junction of the superconducting qubit. (b) Energy level structure of the qutrit is shown above. The lowest three energy levels
of the qutrit are labeled as $|g\rangle_q$, $|e\rangle_q$, and $|f\rangle_q$, respectively.}\label{moxingtu}
\end{figure}

In this section, we introduce the physical model for realizing binomial-code geometric quantum computation. As shown in Fig.~\ref{moxingtu}(a), the system contains a three-level qutrit (denoted by $Q_0$) and a 3D superconducting microwave cavity (denoted by $C_0$). The energy level structure of the qutrit is shown in Fig.~\ref{moxingtu}(b). The energy level of the qutrit $Q_0$ is labeled as $|j\rangle_q$ $(j=g,e,f)$. The frequency of the transition $|j\rangle_q\leftrightarrow|j'\rangle_q$ $(j,j'=g,e,f)$ is denoted by $\omega_{j,j'}$, and the frequency of the cavity $C_0$ is represented by $\omega_0$. Initially, the qutrit is in the ground state $|g\rangle_q$, while the cavity $C_0$ is in the vacuum state.

The transition $\ket{g}_q\leftrightarrow\ket{e}_q$ is driven by a classical field, which can be decomposed as three components with the Rabi frequencies $\widetilde{\Omega}_0$, $\widetilde{\Omega}_2$, $\widetilde{\Omega}_4$ and the detunings $\Delta'_0$, $\Delta'_2$, $\Delta'_4$. respectively. The transition $\ket{e}_q\leftrightarrow\ket{f}_q$ is off-resonantly driven by a classical field with the Rabi frequency $\Omega$ and the detuning $\Delta$. Meanwhile, the transition $\ket{e}_q\leftrightarrow\ket{f}_q$ is coupled to the cavity $C_0$ with the coupling strength $\lambda$ and the detuning $(\Delta-\delta)$. The Hamiltonian of the system under the rotating-wave approximation reads
\begin{align}\label{e1}
H&=H_0+H_1+H_\delta+H_\Delta,\cr
H_0&=\sum\limits_{k=0}^{2}\widetilde{\Omega}_{2k}e^{-i\Delta'_{2k}t}|g\rangle_q\langle e|+\text{H.c.}, \cr
H_1&=(\Omega+\lambda{a^\dag})|e\rangle_q\langle f|
+\text{H.c.},\cr
H_\delta&=\delta{a}^\dag{a},\cr
H_\Delta&=-\Delta|f\rangle_q\langle f|,
\end{align}
where $a$ ($a^{\dag}$) is the annihilation (creation) operator for the cavity $C_0$. For simplicity, we consider that $\Omega$ and $\lambda$ are both real in the following discussions.

Giving a small dimensionless quantity $\varsigma\ll1$ \cite{Leghtas2015}, we place ourselves in the regime where
\begin{align}\label{assp}
&\frac{\Omega}{\Delta}\sim\frac{\lambda\sqrt{N+1}}{\Delta}\sim\varsigma,\cr\
&\frac{\Delta_{2k}'}{\Delta}\sim\frac{\delta}{\Delta}\sim\varsigma^2,\cr\
&\frac{\widetilde{\Omega}_{2k}}{\Delta}\sim\varsigma^3,
\end{align}
with the average photon number $N$ in the operation. Under such assumptions, $H_\Delta$ can be considered as the unperturbed part of the Hamiltonian, while $V=H_0+H_1+H_\delta$ can be deemed as a perturbation. The unperturbed eigenstates are $\ket{g}_q$, $\ket{e}_q$, and $\ket{f}_q$ with eigenvalues $E_g=0$, $E_e=0$, and $E_f=-\Delta$, respectively. Here, $\ket{g}_q$ and $\ket{e}_q$ are two degenerate unperturbed eigenstates. Define the projection operator $P=|g\rangle_q\langle g|+|e\rangle_q\langle e|$. According to the degenerate perturbation theory \cite{10.1143/PTP.70.439}, the effective Hamiltonian is given by
\begin{align}\label{e2}
H_{\mathrm{eff}}&=H_\Delta+PVP+PVQ(E_g-H_\Delta)^{-1}QVP\cr
&+PVQ(E_g-H_\Delta)^{-1}QVQ(E_g-H_\Delta)^{-1}QVP+...\cr
&=H_0+H_1'+H_\delta'+H_\Delta+\mathcal{O}(\varsigma^4),
\end{align}
with  $Q=1-P$ and
\begin{align}
H_1'&=\frac{1}{\Delta}[\lambda^2{a^\dag}a+\Omega\lambda(a^\dag+a)+\Omega^2]|e\rangle_q\langle e|,\cr
H_\delta'&=PH_\delta{P}=\delta{a^\dag}a(|g\rangle_q\langle g|+|e\rangle_q\langle e|).
\end{align}
Here, $(E_g-H_\Delta)^{-1}$ is only required to exist in the projection subspace of $Q$. $\mathcal{O}(\varsigma^4)$ are the terms with orders equal to or greater than $\varsigma^4$. When the state $\ket{f}_q$ is initially unoccupied, the term $H_\Delta$ can be omitted in the intermediate effective Hamiltonian $H_{\mathrm{eff}}$ in Eq.~\eqref{e2}.

To further simplify the dynamical analysis, we attempt to diagonalize the term $\widetilde{H}=H_1'+H_\delta'$ in the intermediate effective Hamiltonian $H_{\mathrm{eff}}$ in Eq.~\eqref{e2}, due to the fact that $H_1'/\Delta\sim{H_\delta'}/\Delta=\mathcal{O}(\varsigma^2)$. This diagonalization can be accomplished by applying a displacement transformation to the cavity field with a displacement of $\alpha_0$, where the required parameters are $\widetilde{\omega}=\delta+\lambda^2/\Delta$ and $\Omega=-\widetilde{\omega}\alpha_0\Delta/\lambda$. In this way, we have
\begin{align}
\widetilde{H}=&\widetilde{\omega}[{a^\dag}a-\alpha_0(a^\dag+a)]|e\rangle_q\langle e|\cr
&+\frac{\Omega^2}{\Delta}|e\rangle_q\langle e|+\delta{a^\dag}a|g\rangle_q\langle g|\cr
=&[\widetilde{\omega}A^\dag A+\frac{\Omega^2}{\Delta}-\widetilde{\omega}\alpha_0^2]|e\rangle_q\langle e|+\delta{a^\dag}a|g\rangle_q\langle g|,
\end{align}
with $A=a-\alpha_0=D(\alpha_0)aD^\dag(\alpha_0)$ and $D(\alpha_0)=\exp[\alpha_0(a^\dag-a)]$. The eigenstates of $A$ are the displacement Fock state $\ket{\widetilde{n}}_c=D(\alpha_0)\ket{n}
_c$. The state $\ket{\widetilde{n}}_c$ can be expressed in the Fock state basis as
\begin{align}
\ket{\widetilde{n}}_c=\sum\limits_{m=0}^\infty\beta_{m,n}\ket{m}_c,
\end{align}
with the expansion coefficients satisfy $\sum_{m=0}^\infty|\beta_{m,n}|^2=1$. Specially, for the displacement vacuum state $\ket{\widetilde{0}}_c$, we have $\beta_{m,0}=\exp(-\alpha_0^2)\alpha_0^m/\sqrt{m!}$.

\begin{widetext}
Then, we expand $H_0$ as
\begin{align}
H_0&=\sum\limits_{k=0}^{2}\widetilde{\Omega}_{2k}e^{-i\Delta'_{2k}t}|g\rangle_q\langle e|+\text{H.c.}=\sum\limits_{k=0}^{2}\widetilde{\Omega}_{2k}e^{-i\Delta'_{2k}t}\left(\sum_{n=0}^\infty|\widetilde{n}\rangle_c\langle\widetilde{n}|\right)|g\rangle_q\langle e|+\text{H.c.}\cr
&=\sum\limits_{k=0}^{2}\sum_{n=0}^\infty\widetilde{\Omega}_{2k}e^{-i\Delta'_{2k}t}|\widetilde{n}\rangle_c\langle\widetilde{n}|\otimes|g\rangle_q\langle e|+\text{H.c.}=\sum\limits_{k=0}^{2}\sum_{m,n=0}^\infty\widetilde{\Omega}_{2k}e^{-i\Delta'_{2k}t}\beta_{m,n}|m\rangle_c\langle \widetilde{n}|\otimes|g\rangle_q\langle e|+\text{H.c.}
\end{align}
Move into a rotation frame via the unitary operator $R=\exp(-i\widetilde{H}t)$. This transforms the effective Hamiltonian $H_{\mathrm{eff}}$ in Eq.~\eqref{e2} into
\begin{align}\label{HR}
H_R=R^\dag{H}_{\mathrm{eff}}R-iR^\dag\dot{R}=\sum\limits_{k=0}^{2}\sum_{m,n=0}^\infty\widetilde{\Omega}_{2k}e^{i(m\delta-n\widetilde{\omega}+\alpha_0^2\widetilde{\omega}-\frac{\Omega^2}{\Delta}-\Delta'_{2k})t}\beta_{m,n}|m\rangle_c\langle \widetilde{n}|\otimes|g\rangle_q\langle e|+\text{H.c.}
\end{align}
\end{widetext}
Considering that the code subspace contains Fock states $\{\ket{0}_c,\ket{2}_c,\ket{4}_c\}$, we select
\begin{align}
\Delta_{2k}'=2k\delta+\alpha_0^2\widetilde{\omega}-\frac{\Omega^2}{\Delta}.
\end{align}
Recalling the relation
\begin{align}
\frac{\widetilde{\omega}}{\Delta}\sim\frac{\delta}{\Delta}\sim\frac{\Omega^2}{\Delta^2}\sim\varsigma^2,\ \ \ \ \ \
\frac{\widetilde{\Omega}_{2k}}{\Delta}\sim\varsigma^3,
\end{align}
we omit the terms with high-frequency oscillations in Eq.~\eqref{HR} and obtain
\begin{align}\label{HR2}
H_R'=&(\widetilde{\Omega}_0\beta_{0,0}|0\rangle_c\langle\widetilde{0}|+\widetilde{\Omega}_2\beta_{2,0}|2\rangle_c\langle\widetilde{0}|\cr
&+\widetilde{\Omega}_4\beta_{4,0}|4\rangle_c\langle\widetilde{0}|)
\otimes|g\rangle_q\langle e|+\mathrm{H.c.}
\end{align}
Finally, substituting
\begin{align}
\widetilde{\Omega}_0&=\Omega_0\cos{\theta}/\beta_{0,0},\cr
\widetilde{\Omega}_2&=\Omega_0\sin{\theta}/\beta_{2,0},\cr
\widetilde{\Omega}_4&=\Omega_0\cos{\theta}/\beta_{4,0},
\end{align}
into $H_R'$ in Eq.~\eqref{HR2}, we obtain the effective Hamiltonian for realizing the binomial-code geometric quantum computation as
\begin{align}\label{eff0}
H_e=\Omega_0\left(\cos\theta|\mathbbmss{O}\rangle_c+\sin\theta|\mathbbm{1}\rangle_c\right)\langle\widetilde{0}|\otimes|g\rangle_q\langle e|+\text{H.c.}
\end{align}

Introducing dressed states
\begin{align}
&|+\rangle_c=\cos\theta|\mathbbmss{O}\rangle_c+\sin\theta|\mathbbm{1}\rangle_c,\cr
&|-\rangle_c=\sin\theta|\mathbbmss{O}\rangle_c-\cos\theta|\mathbbm{1}\rangle_c,
\end{align}
the effective Hamiltonian in Eq.~\eqref{eff0} can be rewritten as
\begin{align}\label{eq12}
H_e&=\Omega_0|+\rangle_c\langle \widetilde{0}|\otimes|g\rangle_q\langle e|+\text{H.c.},\cr\cr
&=\frac{1}{2}\left[\Omega_x(t)\sigma_x+\Omega_y(t)\sigma_y\right]+0\times\sigma_z,
\end{align}
with $\Omega_0=[\Omega_x(t)+i\Omega_y(t)]/2$, $|+,g\rangle=|+\rangle_c|g\rangle_q$, and $|\widetilde{0},e\rangle=|\widetilde{0}\rangle_c|e\rangle_q$, and effective Pauli operators
\begin{align}
\sigma_x&=|+,g\rangle\langle \widetilde{0},e|+|\widetilde{0},e\rangle\langle+,g|,\cr\cr
\sigma_y&=-i|\widetilde{0},e\rangle\langle+,g|+i|+,g\rangle\langle \widetilde{0},e|,\cr\cr
\sigma_z&=|\widetilde{0},e\rangle\langle\widetilde{0},e|-|+,g\rangle\langle +,g|.
\end{align}

\section{NONADIABATIC GEOMETRIC QUANTUM COMPUTATION OF BINOMIAL CODE}\label{III}
\subsection{Evolution path construction via invariant-based reverse engineering}


To study the evolution governed by the two-level effective Hamiltonian $H_e$ in Eq.~(\ref{eq12}) via invariant-based reverse engineering \cite{PhysRevA.83.062116,PhysRevA.86.033405,PhysRevA.109.062610}, we first 
find a dynamical invariant $I_g(t)$ fulfilling \cite{LewisJMP10_1969}
\begin{eqnarray}\label{eq15}
i\frac{\partial}{\partial t}I_g(t)-[H_e, I_g(t)]=0.
\end{eqnarray}
Since the Hamiltonian $H_e$ in Eq.~(\ref{eq12}) owns an SU(2) dynamical structure, a dynamical invariant is found as \cite{PhysRevA.89.043408}
\begin{eqnarray}
I_g(t)&=&\sin\gamma_1(t)\sin\gamma_2(t)\sigma_x+\sin\gamma_1(t)\cos\gamma_2(t)\sigma_y\cr\cr
&&+\cos\gamma_1(t)\sigma_z,
\end{eqnarray}
where $\gamma_1(t)$ and $\gamma_2(t)$ are two time-dependent parameters.
In the following, without causing confusion, $\gamma_1(t)$ and $\gamma_2(t)$ are shorten by $\gamma_1$ and $\gamma_2$. By reversely solving Eq.~(\ref{eq15}), the expressions of $\Omega_x(t)$ and $\Omega_y(t)$ are given by
\begin{eqnarray}
\Omega_x(t)&=&-\left(\dot{\gamma_1}\cos\gamma_2-\dot{\gamma_2}\tan\gamma_1\sin\gamma_2\right)/2,\cr\cr
\Omega_y(t)&=&\left(\dot{\gamma_1}\sin\gamma_2+\dot{\gamma_2}\tan\gamma_1\cos\gamma_2\right)/2.
\end{eqnarray}
The eigenvectors of the dynamic invariant $I_g(t)$ can also be obtained as
\begin{eqnarray}\label{eq21}
\vert\phi_+(t)\rangle&=&\cos\frac{\gamma_1}{2}\vert\widetilde{0},e\rangle+ie^{-i\gamma_2}\sin\frac{\gamma_1}{2}\vert+,g\rangle, \cr\cr
\vert\phi_-(t)\rangle&=&ie^{i\gamma_2}\sin\frac{\gamma_1}{2}\vert\widetilde{0},e\rangle+\cos\frac{\gamma_1}{2}\vert+,g\rangle,
\end{eqnarray}
with the eigenvalues $1$ and $-1$, respectively. Using the eigenvectors $|\phi_{\pm}(t)\rangle$, two orthogonal solutions $|\Psi_\pm(t)\rangle$ of the Schr\"{o}dinger equation $i|\dot{\Psi}(t)\rangle=H_e|\Psi(t)\rangle$ are derived as
\begin{eqnarray}
|\Psi_\pm(t)\rangle&=&e^{i\mu_\pm(t)}\vert\phi_\pm(t)\rangle,
\end{eqnarray}
where $\mu_{\pm}(t)$ are the Lewis-Riesenfeld (LR) phases given by \cite{LewisJMP10_1969}
\begin{align}\label{eq19}
\mu_{\pm}(t)=\theta_d^{\pm}+\theta_g^{\pm},
\end{align}
with the dynamic part
\begin{align}
 \theta_d^{\pm}=-\int_0^t\langle\phi_{\pm}(t')|H_e|\phi_{\pm}(t')\rangle dt',
\end{align}
and the geometric part
\begin{align}
\theta_g^{\pm}=\int_0^t\langle\phi_{\pm}(t')|i\frac{\partial}{\partial t'}|\phi_{\pm}(t)\rangle dt',
\end{align}
respectively.
According to the results of Eqs.~ (\ref{eq21}) and (\ref{eq19}), the time derivatives of the dynamic phases and the geometric phases are acquired by $|\phi_{\pm}(t)\rangle$ as 
\begin{eqnarray}
\dot{\theta}_d^{\pm}(t)=\mp\frac{\dot{\gamma}_2\sin^2\gamma_1}{2\cos\gamma_1},\ \ \ \dot{\theta}_g^{\pm}(t)=\pm\dot{\gamma}_2\sin^2\frac{\gamma_1}{2}.
\end{eqnarray}

To achieve 
geometric quantum computation base on binomal code, it is necessary
to eliminate the dynamic part of the Lewis-Riesenfeld phase
obtained in [0,$T$] as
\begin{eqnarray}
\theta_d^{\pm}&=&-\int_0^t\langle\phi_{\pm}(t')|H_e|\phi_{\pm}(t')\rangle dt'=0,
\end{eqnarray}
where $T$ is the total evolution time. In this case, the remaining part of Lewis-Riesenfeld phase,
\begin{eqnarray}
\theta_g^{\pm}&=&\int_0^t\langle\phi_{\pm}(t')|i\frac{\partial}{\partial t'}|\phi_{\pm}(t)\rangle dt',
\end{eqnarray}
where is purely geometric phase. Moreover, in the evolution process, the cyclic evolution condition also needs to be satisfied, which are described by
\begin{eqnarray}
&|\phi_{\pm}(0)\rangle=|\phi_{\pm}(T)\rangle\Rightarrow\gamma_1(0)=\gamma_1(T)=0.
\end{eqnarray}
Moreover, $\gamma_1(t)$ should be symmetrical about $T/2$, i.e., $\gamma_1(t)=\gamma_1(T-t)$.
Based on the above conditions, the time-dependent parameters can be designed as
\begin{eqnarray}\label{eqs22}
\gamma_2(t)&=&-\Theta_g\xi(t)+\widetilde{\gamma}_2(t),\cr\cr
\xi(t)&=&\begin{cases}
0,\ \ [0,T/2),\cr\cr
1,\ \ [T/2,T],
\end{cases}
\end{eqnarray}
where $\widetilde{\gamma}_2(t)$ is an undetermined analytic function and symmetrical about $T/2$, i.e., $\widetilde{\gamma}_2(t)=\widetilde{\gamma}_2(T-t)$. Under the above parameter design, pure geometric phase $\theta_g^{-}(T)=\Theta_g$ can be achieved and the dynamic phase $\theta_d^-(T)$ is completely eliminated after the evolution (detail in the appendix \ref{appenA}), where $\vert\phi_-(t)\rangle$ is chosen as the evolution path. Thus, the evolution operator of the system after the whole operation can be described by $U(T)=U_g(T)\otimes|g\rangle_q\langle g|$ with
\begin{align}
U_g(T)=e^{i\Theta_g}|+\rangle_c\langle+|+|-\rangle_c\langle-|.
\end{align}
In the computational subspace $\mathcal{S}_c=\mathrm{span}\{\ket{\mathbbmss{O}}_c,\ket{\mathbbm{1}}_c\}$, it can be displayed in a matrix form
\begin{align}\label{Ug}
U_g(T)=\begin{bmatrix}
e^{i\Theta_g}\cos^2\theta-\sin^2\theta & \sin\theta\cos\theta\left(e^{i\Theta_g}-1\right) \\[2ex]
\sin\theta\cos\theta\left(e^{i\Theta_g}-1\right) & e^{i\Theta_g}\sin^2\theta+\cos^2\theta
\end{bmatrix}.
\end{align}

According to the operator $U_g(T,0)$ in Eq.~\eqref{Ug}, we can realize arbitrary single-qubit gates by selecting different parameters $\theta$ and $\Theta_g$. The values of $\theta$ and $\Theta_g$ for realizing serval useful single-qubit gates are listed in Table~\ref{tab1}.

\begin{table}\centering
\caption{Parameters of different quantum gates.}\label{tab1}
{\begin{tabular}{ccccc} \hline\hline
&&\\[-7pt]
\ \ Gates\ & \ $\theta$ \ & \ \ $\Theta_g$  \ & \ \ $\gamma_1(0)$ [$\gamma_1(T)$] \  & \ $\gamma_2(0)$ [$\gamma_2(T)$]\  \\[1pt]
\hline
&&\\[-7pt]
\ \ $\pi$-phase \ \ &\ \ $\pi$\ & \ $\pi$\ \   & \ $0$ [0] \  & \ $0$ [$\pi$]  \ \   \\
&&\\[-4pt]
\ \ NOT \ \ &\ \ \ $\pi/4$\ & \ \ $\pi$\ \ \ & \ $0$ [0] \  & \ $0$ [$\pi$]  \ \  \\
&&\\[-4pt]
\ \ Hadamard \ \ &\ \ $\pi/8$\ & \ \ $\pi$\ \ \ & \ $0$  [0] \  &\ $0$ [$\pi$]  \ \
\\[2pt]
\hline \hline
\end{tabular}}
\end{table}

\subsection{Parameter design via optimal control theory}

Systematic error, as a common error, is widely prevalent in actual experiments, which describes many low-order processes of complex errors and many imperfections of instruments and operations \cite{Ruschhaupt_2012,PhysRevLett.111.050404,KYHPRA111_2025}. In order to find an evolution path resistant to the systematic error, we consider the optimal control method with the nullification of systematic-error sensitivity. Specifically, when the control field is affected by systematic error with error rate $\varepsilon$, i.e., $\widetilde{\Omega}_{2k}\rightarrow(1+\varepsilon)\widetilde{\Omega}_{2k}$ ($k=0,1,2$) , the effective Hamiltonian $H_e$ in Eq. (\ref{eq12}) becomes $(1+\varepsilon)H_e$.
By using the time-dependent perturbation theory, one gets \cite{Ruschhaupt_2012}
\begin{eqnarray}
|\Psi_\pm^{\varepsilon}(t)\rangle&=&|\Psi_\pm(T)\rangle-i\varepsilon\int_0^{T}dtU_g(T,t)H_e|\Psi_\pm(t)\rangle\cr&+&\mathcal{O}(\varepsilon^2),
\end{eqnarray}
where $U_g(T,t)=\sum_{l=\pm}|\Psi_\pm(T)\rangle\langle\Psi_\pm(t)|$ is the unperturbed time evolution operator for the time interval $[t,T]$, $\mathcal{O}(\varepsilon^2)$ are the terms with orders equal to or greater than $\varepsilon^2$, and $|\Psi_\pm^{\varepsilon}(t)\rangle$ ($|\Psi_\pm(t)\rangle$) is the state of the system with (without) the influence of the systematic errors. Since the state $|-\rangle$ is dynamically decoupled from the Hamiltonian $H_e$ and the cycling evolution is governed by path $|\Psi_-(t)\rangle=e^{i\mu_-(t)}\vert\phi_-^s(t)\rangle$ with the initial state $|+,0\rangle$, the fidelity of the evolution can be approximately calculated as
\begin{eqnarray}
F_g&=&1-\varepsilon^2\left|\int_0^{T}e^{2i\mu_-(t)}\langle\phi_+^s(t)\vert H_e\vert\phi_-^s(t)\rangle dt\right|^2\cr&+&O(\varepsilon^3),
\end{eqnarray}
by considering the relations $\mu_-(t)=-\mu_+(t)$ between the LR phases acquired in two different paths. The systematic error sensitivity $Q_g$ is given by \cite{Ruschhaupt_2012}
\begin{eqnarray}
Q_g=-\frac{1}{2}\frac{\partial^2F_g}{\partial\varepsilon^2}\mid_{\varepsilon=0}
=\left|\int_0^{T}e^{i\chi(t)}\dot{\gamma_1}\sin^2\gamma_1 dt\right|^2,
\end{eqnarray}
with $\chi(t)=\gamma_2+2\mu_-(t)$.

\begin{figure}
  \centering
  \includegraphics[width=\linewidth]{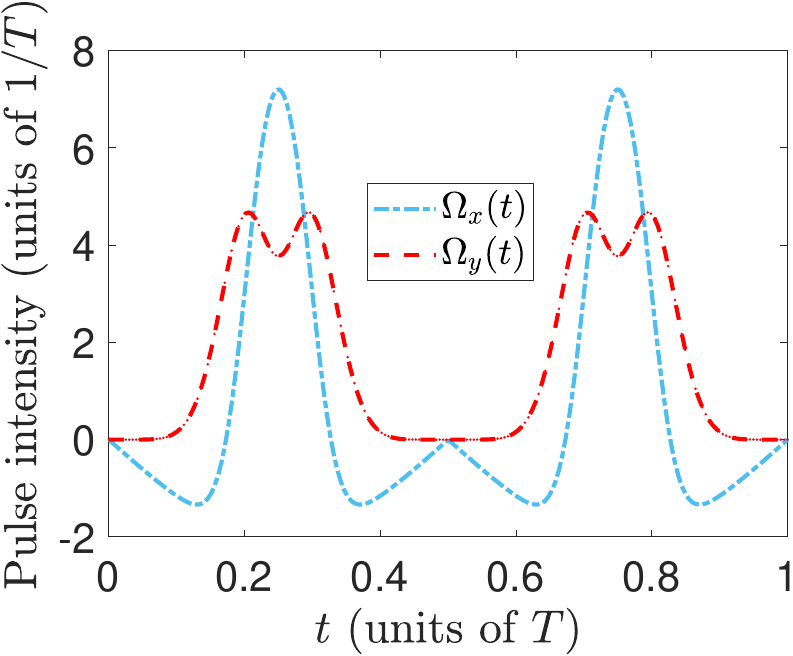}\caption{Control fields $\Omega_x(t)$ and $\Omega_y(t)$ versus $t/T$.}\label{boxingtu}
\end{figure}

\begin{figure}
  \centering
  \includegraphics[width=8cm]{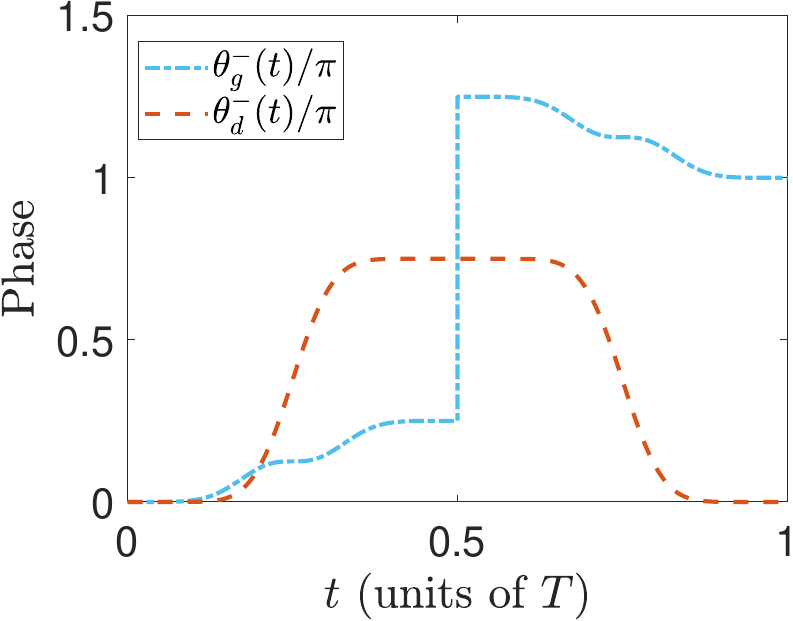}\caption{Dynamic phase $\theta_d^-(t)$ and geometric phase $\theta_g^-(t)$ versus $t$ in the implementation of geometric quantum gates.}\label{dynamicphase}
\end{figure}

To nullify the systematic-error sensitivity, we select
\begin{align}
\chi(t)=\Theta_g\xi(t)+\chi_0\left[2\gamma_1(t)-2\sin[2\gamma_1(t)]\right],
\end{align}
with a time-independent coefficient $\chi_0$, 
yielding
\begin{eqnarray}
Q_g=\sin^2(\chi_0\pi)\sin(\Theta_g^2/2)/\chi_0^2.
\end{eqnarray}
This implies that $Q_g=0$ can be obtained with $\chi_0=\pm1,\pm2,\cdots$. Here, $\chi_0=1$ is chosen for simplicity. Using the relation $\chi(t)=\gamma_2+2\mu_-(t)$, one derives
\begin{eqnarray}
\dot{\chi}(t)&=&2\dot{\gamma_1}(1-\cos2\gamma_1)=\dot{\gamma_2}/\cos\gamma_1,\cr\cr
\Rightarrow\dot{\gamma_2}(t)&=&2\dot{\gamma_1}\cos\gamma_1(1-\cos2\gamma_1).
\end{eqnarray}
According to the boundary conditions in Table \ref{tab1}, the expression of $\gamma_2$ is given by
\begin{eqnarray}\label{e18}
\gamma_2(t)&=&-\Theta_g\xi(t)+\frac{4}{3}\sin^3\gamma_1.
\end{eqnarray}
In addition, based on the boundary conditions in Table \ref{tab1}, the parameter $\gamma_1(t)$ can be selected as
\begin{eqnarray}
\gamma_1(t)&=&\pi\sin^2(\pi t/T).
\end{eqnarray}
At this point, the wave shapes of $\Omega_x(t)$ and $\Omega_y(t)$ can be obtained and are shown in Fig.~\ref{boxingtu}. Based on the above parameter design, the dynamic phase $\theta_g^-(t)$ and geometric phase $\theta_g^+(t)$ versus $t$ are also investigated, which are shown in Fig. \ref{dynamicphase}. The results show that the dynamic phase is completely eliminated and the desired geometric phase is obtained after the evolution.

\section{NUMERICAL SIMULATIONS AND DISCUSSIONS}\label{IV}

In this section, numerical simulations are performed to check the feasibility and robustness of the protocol. To meet the assumptions in Eq.~\eqref{assp}, parameters $T=5~\mu$s, $\delta=-2\pi\times12$~MHz, $\lambda=2\pi\times462$~MHz, and $\Delta=2\pi\times4.78$~GHz are chosen. Here, we consider $Q_0$ as a flux qutrit with the transition frequencies of $\ket{g}_q\leftrightarrow\ket{e}_q$ and $\ket{e}_q\leftrightarrow\ket{f}_q$ as $2\pi\times3$~GHz, and $2\pi\times12$~GHz, respectively \cite{PhysRevA.105.042434,PhysRevB.86.140508,PhysRevLett.115.223603}. The frequency of cavity $C_0$ is $\omega_0=2\pi\times16.792$ GHz. To avoid extremely small coefficients $\beta_{2k,0}$ which leads to large Rabi frequency $\widetilde{\Omega}_{2k}$, the parameter $\alpha_0$ is selected as $\alpha_0=\sqrt{2}$. In this way, the average photon number of the intermediate state $\ket{\widetilde{0}}_c$ is $\braket{\widetilde{0}|a^\dag{a}|\widetilde{0}}_c=\alpha_0^2=2$, equal to the average photon numbers of the logical states $\ket{\mathbbmss{O}}_c$ and $\ket{\mathbbm{1}}_c$ of the binomial code. The expansion parameters take the values $\beta_{0,0}=0.1353$, $\beta_{2,0}=0.1914$, and $\beta_{4,0}=0.1105$. The feasibility and robustness of the protocol are discussed in detail in three parts below.

\subsection{Verification of the protocol with different quantum gates}

To verify the feasibility of the protocol, we investigated the evolution operator of system under the original Hamiltonian $H$ in Eq.~(\ref{e1}) and the effective Hamiltonian $H_e$ in Eq.~(\ref{eq12}). Here, to evaluate the performance of geometric quantum gates, the average fidelity of a gate over all possible initial states is defined as
\begin{eqnarray}
\overline{F}_g(t)=\frac{1}{l(l+1)}\left\{\mathrm{Tr}\left[M^{\dag}M\right]+ \left|\mathrm{Tr}\left[M\right]\right|^2\right\},
\end{eqnarray}
with $M=\mathcal{P}_cU_T^{\dag}U(t)\mathcal{P}_c$, the dimension $l=2$ and the projection operator $\mathcal{P}_c=|\mathbbmss{O}\rangle_c\langle\mathbbmss{O}|+|\mathbbm{1}\rangle\langle_c\mathbbm{1}|\otimes|g\rangle_q\langle g|$ of the computational subspace. Here, the target operator $U_T$ is respectively set as $U_{\pi}$ ($\pi$-phase gate), $U_{\mathrm{NOT}}$ (NOT gate), and $U_{\mathrm{Hadamard}}$ (Hadamard gate) to examine the feasibility of the protocol, with
\begin{align}
&U_{\pi}=\left[|\mathbbmss{O}\rangle_c\langle\mathbbmss{O}|-|\mathbbm{1}\rangle_c\langle\mathbbm{1}|\right]\otimes|g\rangle_q\langle g|,\cr
&U_{\mathrm{NOT}}=\left[|\mathbbmss{O}\rangle_c\langle\mathbbm{1}|+|\mathbbm{1}\rangle_c\langle\mathbbmss{O}|\right]\otimes|g\rangle_q\langle g|,\cr
&U_{\mathrm{Hadamard}}=\frac{1}{\sqrt{2}}(U_{\pi}+U_{\mathrm{NOT}}).
\end{align}

\begin{figure}
  \centering
  \subfigure{\scalebox{0.3}{\includegraphics{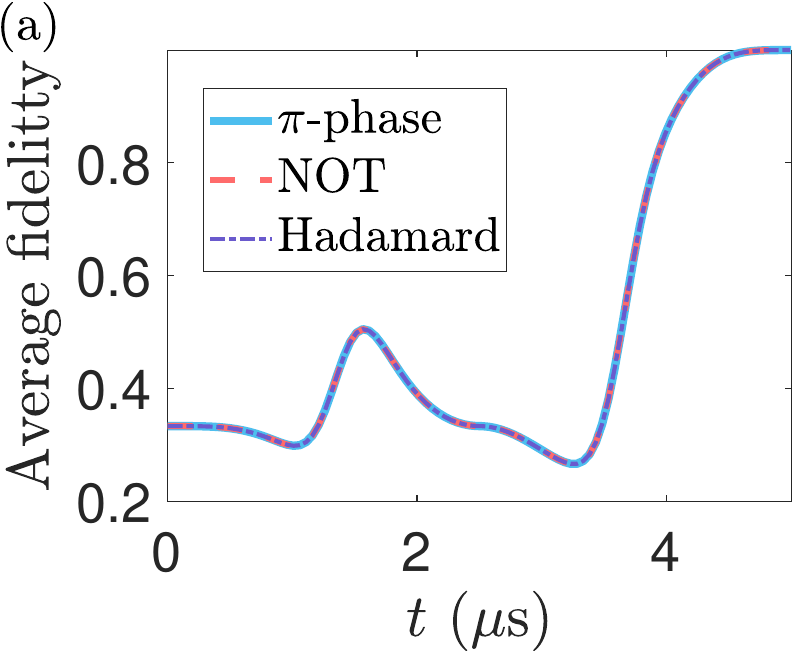}} \label{effthreegate}}
  \subfigure{\scalebox{0.3}{\includegraphics{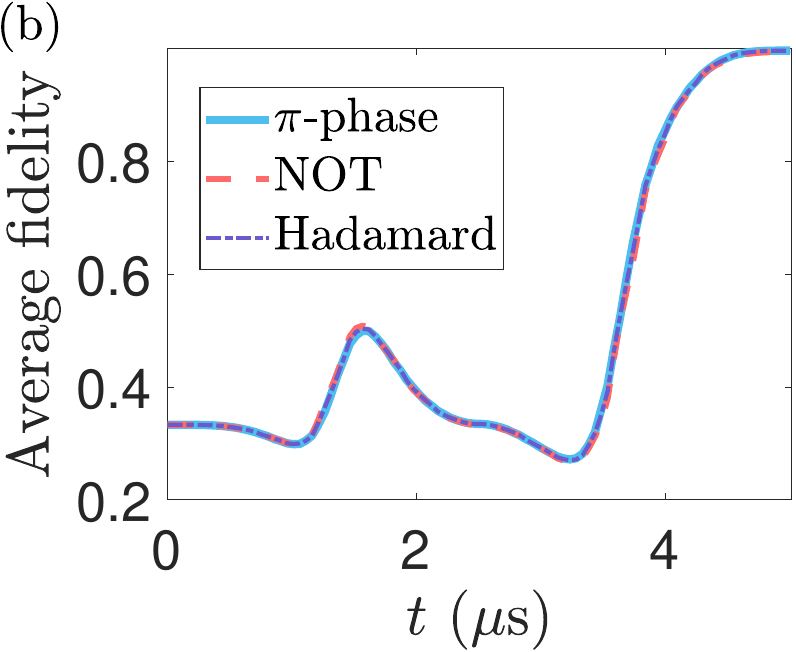}}\label{threegate} }
  \caption{Average fidelity $F_g(t)$ versus $t$ for different gates, where the system is governed by (a) the effective Hamiltonian $H_e$ and (b) 
  the full Hamiltonian $H$.}\label{fig3}
\end{figure}

As shown in Figs \ref{effthreegate} and \ref{threegate}, the average fidelities $\overline{F}_g(t)$ of three quantum gates versus $t$ are plotted when the system is controlled by the effective Hamiltonian $H_e$ in Eq.~(\ref{eq12}) and the full Hamiltonian $H$ in Eq.~(\ref{e1}), respectively. Apparently, when the system is controlled by the effective Hamiltonian $H_e$, we can precisely obtain the target geometric quantum gates, i.e., the average fidelity $\overline{F}_g(T)$ can reach 1 at final moment. This confirms the effectiveness of pulse design. Moreover, the average fidelity $\overline{F}_g(T)$ is over 0.99 when the evolution governed by the full Hamiltonian $H$ in Eq.~(\ref{e1}) is completed for different geometric quantum gates. This indicates that, the real system evolution governed by the full Hamiltonian $H$ in Eq.~(\ref{e1}) is consistent with the evolution driven by the effective Hamiltonian $H_e$ in Eq.~(\ref{eq12}) under reasonable parameter selection and the desired quantum gates can be obtained after the evolution.

\subsection{Effects of systematic errors on the protocol}\label{Ab}

\begin{figure}
  \centering
  \includegraphics[width=\linewidth]{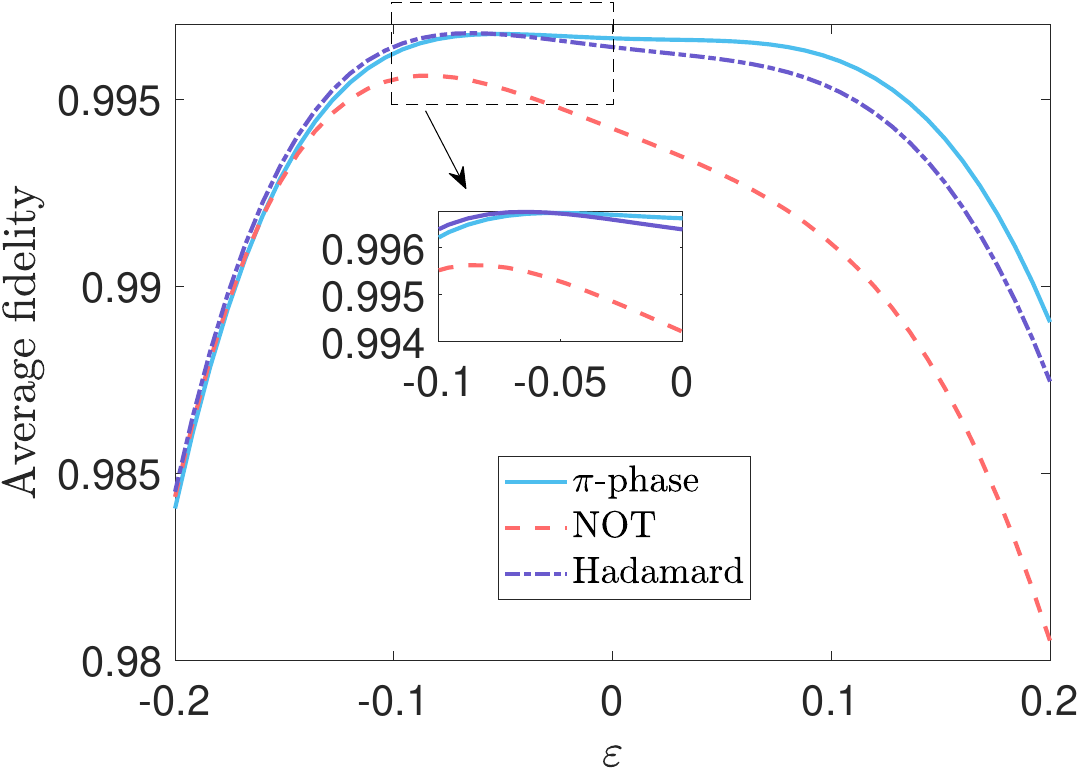}\caption{Final average fidelity $\overline{F}_g(T)$ versus the systematic error rate $\varepsilon$ for different quantum gates.}\label{epsilont}
\end{figure}

We now consider thee influence of the systematic errors in the control parameters. In the present protocol, there may exist systematic errors in the strength of the Rabi frequency $\widetilde{\Omega}_{2k}$, i.e., $\widetilde{\Omega}_{2k}\rightarrow(1+\varepsilon)\widetilde{\Omega}_{2k}$ ($k=0,1,2$). Here, $\varepsilon$ denotes the error rate. The final average fidelity $\overline{F}_g(T)$ versus the error rate $\varepsilon$ is plotted in Fig.~\ref{epsilont} for different quantum gates. On the whole, the final average fidelity $\overline{F}_g(T)$ decreases as $|\varepsilon|$ increases for the error rate $\varepsilon\in[-0.2,0.2]$ \cite{PhysRevA.107.013702,PhysRevA.109.042615,Zhao2021}. However, the value of $\overline{F}_g(T)$ is always greater than $99\%$ when $\varepsilon\in[-0.2,0.2]$ for three kinds of quantum gates. This demonstrates that the control approach given in Sec.~\ref{III} 
makes the protocol robust to the systematic errors of the Rabi frequency $\widetilde{\Omega}_{2k}$.

Moreover, as shown in Fig.~\ref{epsilont}, the maximum final average fidelities $\overline{F}_g(T)$ for all three quantum gates do not occur at zero systematic error ($\varepsilon=0$). Instead, it is achieved within a small negative error range of $-0.1<\varepsilon<0$. This result arises from competing effects. While a negative shift in the Rabi frequency ($\varepsilon<0$) inherently degrades its own precision, it simultaneously better satisfies the condition $\widetilde{\omega}\gg \widetilde{\Omega}_{2k}$. This enhances the accuracy of the  effective dynamics of the system. Consequently, the accuracy gain from the more valid effective dynamics can outweigh the detrimental impact of the Rabi frequency shift. This net positive effect explains why a slight negative deviation in $\varepsilon$ leads to peak fidelity for these gates.

\subsection{Effects of additive white Gaussian noise on the protocol}

\begin{figure}
\centering
\subfigure{\scalebox{0.23}{\includegraphics{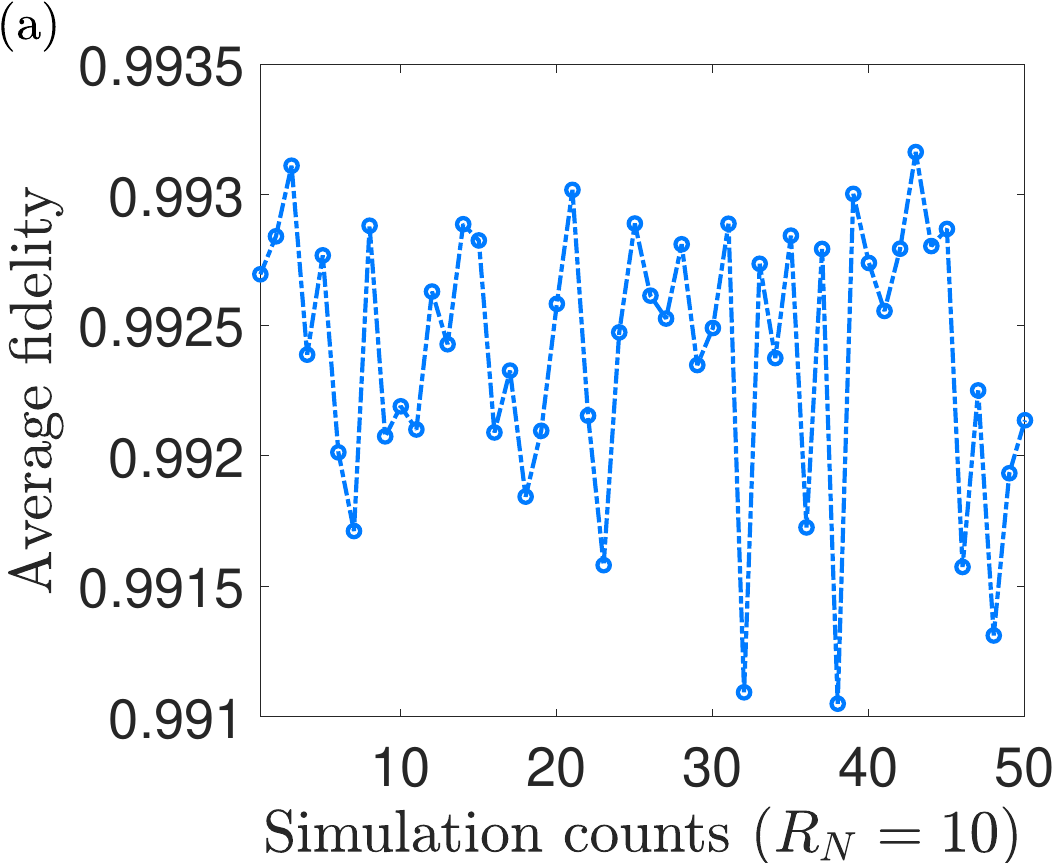}} \label{simulationcount}}
\subfigure{\scalebox{0.23}{\includegraphics{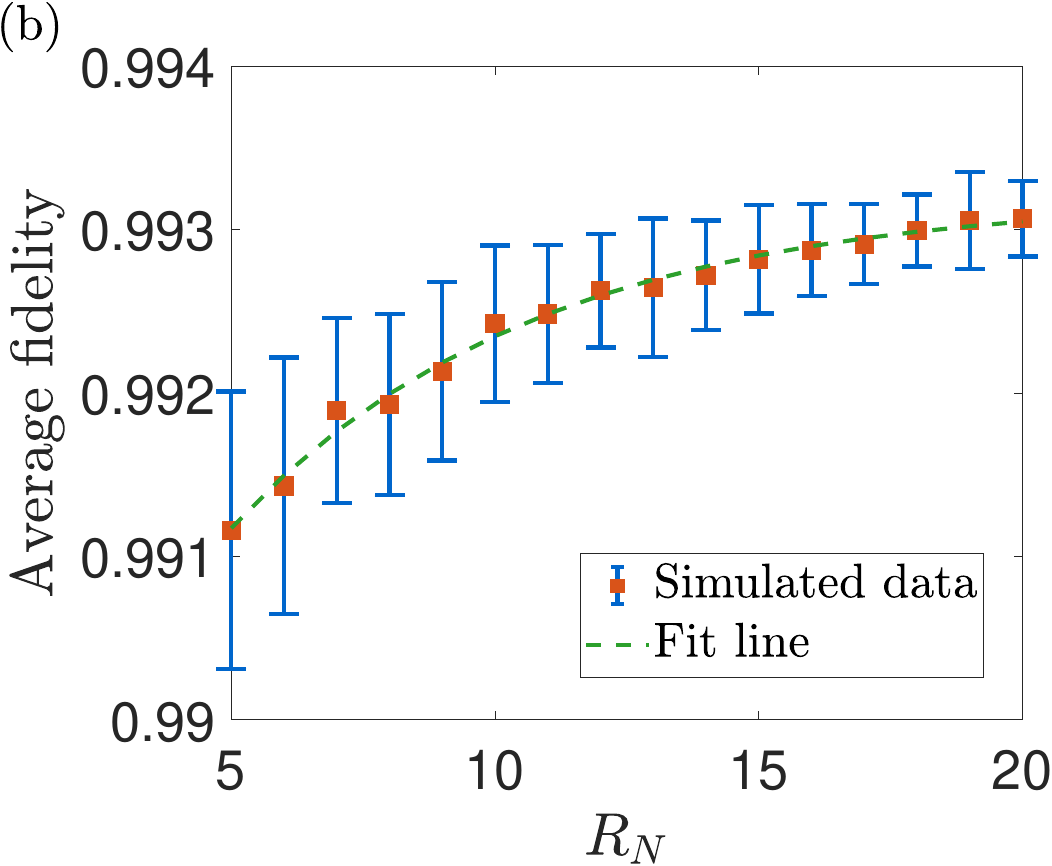}}\label{snr2} }
\caption{(a) Final average fidelity $\overline{F}_g(T)$ versus simulation counts under AWGN with SNR $R_N=10$. (b) Joint average fidelity $\mathcal{\overline{F}}_g(T)$ (averages over 50 samples and all possible initial states) versus SNR $R_N$. The green dashed line exhibits the fitting results based on $\mathcal{\overline{F}}_g(T)=a\times\exp(-b\times{R_N})+c$, where $a=-0.004824$, $b=0.1734$, and $c=0.9932$.}\label{fig5}
\end{figure}

In addition to systematic errors, random noise in the experimental environment inevitably affects the average fidelity of quantum gates. Given that additive white Gaussian noise (AWGN) serves as an effective model for emulating a variety of stochastic processes, we evaluate the performance of the protocol by introducing AWGN into the control fields. Consequently, the control fields take the form
\begin{eqnarray}
\Omega_x^{\mathrm{AWGN}}(t)&=&\Omega_x(t)+\mathrm{awgn}\left[\Omega_x(t),R_N\right],\cr\cr
\Omega_y^{\mathrm{AWGN}}(t)&=&\Omega_y(t)+\mathrm{awgn}\left[\Omega_y(t),R_N\right],
\end{eqnarray}
where $\mathrm{awgn}\left[\Omega_x(t),R_N\right]$ ($\mathrm{awgn}\left[\Omega_y(t),R_N\right]$) is a function that generates AWGN mixed into the control fields $\Omega_x(t)$ [$\Omega_y(t)$] with a signal-to-noise ratio (SNR) $R_N$. Unlike the systematic errors, owing to the distinct effects of random noise in two independent simulations, multiple numerical simulations are required to evaluate the impact of AWGN. Here, 
we 
take the target operation $U_T=U_{\mathrm{NOT}}$ as an example. 
Fifty simulations are performed for the final average fidelity $\overline{F}_g(T)$ with SNR $R_N=10$, and the results are shown in Fig.~\ref{simulationcount}.  

As shown in Fig.~\ref{simulationcount}, the final average fidelity $\overline{F}_g(T)$ exhibits only minor fluctuations ($<0.003$) in the presence of AWGN. In all 50 simulations, the final average fidelity $\overline{F}_g(T)$ remain above 0.99, indicating the 
robustness of the protocol against AWGN. To further investigate the influence of random noise, we consider the joint average fidelity $\mathcal{\overline{F}}_g(T)$, averaging over 50 samples and all possible initial states per point of SNR with $R_N\in[5,20]$. 
The results are shown in Fig.~\ref{snr2}, where each error bar is obtained by calculating standard deviation in 50 samples with corresponding SNR. 
Through curve fitting, we find that the relationship between the joint average fidelity and SNR is given by $\mathcal{\overline{F}}_g(T)=a\times\exp(-b\times{R_N})+c$, where $a=-0.004824$, $b=0.1734$, and $c=0.9932$. Clearly, the joint average fidelity consistently exceeds 0.99 for $R_N\in[5,20]$. This indicates that the protocol exhibits strong robustness against random noise.

\subsection{Effects of decoherence on the protocol}\label{Ac}

Given that the system cannot be fully isolated from the external environment, the decoherence induced by environmental factors should also be taken into consideration \cite{PhysRevA.84.043832,PhysRevLett.110.243601}. The main decoherence factors in the protocol are the qutrit dephasing on the levels $|f\rangle_q$,
and $|e\rangle_q$, the qutrit energy relaxation of the paths $|f\rangle_q\rightarrow|e\rangle_q$, $|f\rangle_q\rightarrow|g\rangle_q$, and $|e\rangle_q\rightarrow|g\rangle_q$, and the single photon loss of cavity. To investigate the impacts of these decoherence factors, the NOT gate is adopted as a representative example, and the initial and target states of the system are $\ket{\mathbbmss{O}}_c\ket{g}_q $ and $\ket{\mathbbm{1}}_c\ket{g}_q $, respectively. The evolution of the system in the presence of decoherence can be described by the Lindblad master equation \cite{10.1063/5.0134394,KYHSCPMA68_2025}
\begin{eqnarray}\label{mas}
\dot{\rho}(t)&=&-i\left[H,\rho(t)\right] + \sum_{p=e}^f\Gamma_{pp}\mathcal{L}[\sigma_{pp}]\rho(t)\cr\cr
&+&\sum_{i=g,e}\Gamma_{f\iota}\mathcal{L}[\sigma_{f\iota}^-]\rho(t)
+\Gamma_{eg}\mathcal{L}[\sigma_{eg}^-]\rho(t)\cr\cr
&+&\kappa\mathcal{L}[a]\rho(t),
\end{eqnarray}
with $\sigma_{pp}=|p\rangle_q\langle p|$ ($f\!=\!e,f$), $\sigma_{f\iota}^-=|f\rangle_q\langle i|$ ($\iota\!=\!g,e$), and $\sigma_{eg}^-=|e\rangle_q\langle g|$. In addition, $\mathcal{L}[O]$ is the Lindblad superoperator satisfies
\begin{eqnarray}
\mathcal{L}[O]\rho(t)=O\rho(t)O^{\dag}-O^{\dag}O\rho(t)/2-\rho(t)O^{\dag}O/2,
\end{eqnarray}
with $O=\sigma_{pp},\sigma_{f\iota}^-,\sigma_{eg}^-,a$.
Here, $\Gamma_{pp}$ is the dephasing rate, $\Gamma_{f\iota}^-$ ($\Gamma_{eg}$) is the energy relaxation rate, and $\kappa$ is the single photon loss rate of cavity $C_0$. For simplicity, we set $\Gamma_{pp}=\Gamma_{d}$, $\Gamma_{f\iota}=\Gamma_{s}/2,\Gamma_{eg}=\Gamma_{s}$, and $\kappa=\Gamma_{\kappa}$ in following numerical simulations.

\begin{figure}
  \centering
  \includegraphics[width=8cm]{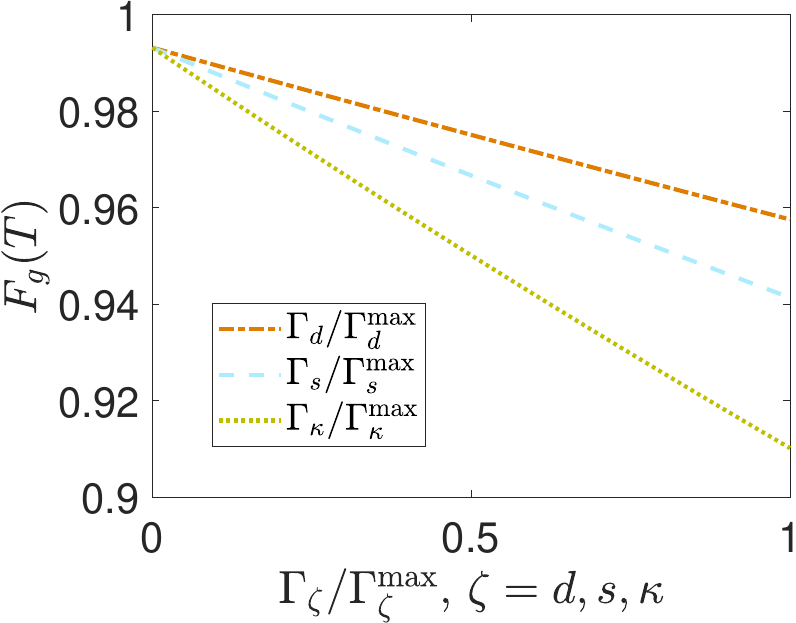}\caption{ Final fidelity $F_g(T)$ versus the dephasing rate $\Gamma_{s}\in[0, 50]$ kHz, the qutrit spontaneous emission rate $\Gamma_{\gamma}\in[0, 50]$ kHz, and the single photon loss rate $\Gamma_{\kappa}\in[0, 10]$ kHz ($\zeta=d,s,\kappa$), where $\Gamma_{d}^{\mathrm{max}}=50$ kHz, $\Gamma_{s}^{\mathrm{max}}=50$ kHz, and $\Gamma_{\kappa}^{\mathrm{max}}=10$ kHz are the maximum value of the dephasing rate $\Gamma_{d}$, the qutrit spontaneous emission rate $\Gamma_{s}$, and the single photon loss rate $\Gamma_{\kappa}$, respectively.}\label{dec}
\end{figure}


Based on the master equation in Eq.~(\ref{mas}), the effects of dephasing, energy relaxation, and single-photon loss are investigated. In Fig.~\ref{dec}, the final fidelity $F_g(T)=\langle g|_q\langle\mathbbm{1}|_c\rho(T)|\mathbbm{1}\rangle_c|g\rangle_q$  versus the dephasing rate $\Gamma_{d}$, the energy relaxation rate $\Gamma_{\gamma}$, and the single photon loss rate $\Gamma_{\kappa}$ of cavity $C_0$ are plotted. As shown in Fig.~\ref{dec}, the qutrit energy relaxation has a greater impact compared with the qutrit dephasing. The orange-dotted-dashed line (blue-dashed line) curve demonstrate that the final fidelity $F_g(T)\geq95\%$ ($F_g(T)\geq94\%$) can be obtained with $\Gamma_{d}$ ($\Gamma_{\gamma}$) in the range $[0,50]$~kHz. 
In addition, Fig.~\ref{dec} shows that the protocol can achieve the final fidelity $F_g(T)\geq 91\%$ for $\Gamma_{\kappa}\in[0,10]$ kHz. This suggests that the single photon loss is the main decoherence factor affecting the protocol, and cavities with higher quality factors are desired.

In superconducting system, the relaxation time $T_1$ and the dephasing time $T_2$ of flux qubits about 40--60~$\upmu$s (corresponding to the relaxation and dephasing rates about 16.67--25~kHz) have been previously reported \cite{Yan2016,5136262}. Accordingly, the considered range $[0,50]$~kHz for the relaxation and dephasing rates in Fig.~\ref{dec} is reasonable and available for flux qubits. Moreover, the high quality factor $Q_c\sim3.5\times10^7$ to $Q_c\sim3\times10^9$ has been demonstrated for 3D microwave cavities \cite{science.aaf2941,PhysRevB.94.014506,PRXQuantum.4.030336}. Substituting the cavity frequency $\omega_0=2\pi\times16.792$ GHz of cavity $C_0$, the ranges of single photon loss rate for the cavity $C_0$ ranges in $[0.035,3.014]$~kHz. Therefore, the maximal single photon loss rate 10~kHz considered in Fig.~\ref{dec} is reasonable. According to the numerical results in this section, the protocol can still realize geometric quantum gates based on binomial code with relatively high fidelity in the presence of decoherence with current technology in superconducting systems.

\section{CONCLUSION}\label{V}

In conclusion, we have presented a noise-resilient protocol for implementing Nonadiabatic geometric quantum computation with binomial codes in a superconducting system comprising a microwave cavity dispersively coupled to a 
qutrit. By appropriately engineering the couplings between the qutrit and the cavity, as well as applying tailored driving fields to the qutrit, 
an effective Hamiltonian is derived. This effective Hamiltonian enables the manipulation of the logical states of binomial codes through an intermediate state.

The control fields 
are designed using reverse engineering and optimal control theory. 
Starting with a Hermitian dynamical invariant operator connected to the effective Hamiltonian, the evolution path for the geometric quantum gate operations is established. Subsequently, optimal control theory directs the optimization of control parameters. Minimizing a cost function, namely the systematic error sensitivity, yields an error-resistant control scheme for nonadiabatic binomial code geometric quantum gates.

Through numerical simulations, we have validated the feasibility of the protocol and examined its robustness against systematic errors in the control fields. Moreover, the effect of random noise, random initial phase, and decoherence on the performance of the protocol have also been considered. The results demonstrate that the protocol can achieve acceptable average fidelity for geometric quantum gates even in the presence of these disturbances. Therefore, the protocol may also be applicable for realizing geometric quantum computation based on binomial code in a wide range of physical systems.


\begin{acknowledgments}
Y.-H.~Kang was supported by the National Key Research and Development Program of China under Grant No.~2024YFA1408900. Y.~Wang was supported by the National Natural Science Foundation of China under Grant No.~12404404. Y.-H.C. was supported by the National Natural Science Foundation of China under Grant No. 12304390 and 12574386, the National Postdoctoral Overseas Talent Recruitment Program of China, the Fujian 100 Talents Program, and the Fujian Minjiang Scholar Program. Y.~Xia was supported by the National Natural Science Foundation of China under Grant No. 62471143, the Key Program of National Natural Science Foundation of Fujian Province under Grant No.~2024J02008. Z.-C.~Shi was supported by the National Natural Science Foundation of China under Grant No.~62571129 and the Natural Science Foundation of Fujian Province under Grant No.~2025J01456.
\end{acknowledgments}

\begin{appendix}
\section{THE DYNAMIC PHASE AND THE GEOMETRIC PHASE OBTAINED FROM THE EVOLUTION PROCESS}\label{appenA}

In this section, we prove that the of dynamic phase can be eliminated and the acquired geometric phase is $\Theta_g$ after the whole operation. According to Eq. (\ref{eqs22}),  $\dot{\gamma}_2(t)$ can be given by $\dot{\gamma}_2(t)=-\Theta_g\delta'(T-2)+\dot{\widetilde{\gamma}}_2(t)$, where $\delta'$ is the Dirac delta function. We assume the variation of $\gamma_1(t)$ and $\widetilde{\gamma}_2(t)$ are symmetrical about $T/2$, i.e., $\gamma_1(t)=\gamma_1(T-t)$ and $\widetilde{\gamma}_2(t)=\widetilde{\gamma}_2(T-t)$.
It is easy to derive
\begin{eqnarray}
\dot{\widetilde{\gamma}}_2(t)=\frac{d\widetilde{t}}{dt}\times\frac{d}{d\widetilde{t}}\widetilde{\gamma}_2(\widetilde{t})=-\widetilde{\gamma}_2(\widetilde{t}),
\end{eqnarray}
with $\tilde{t}=T-t$. Then, the dynamic phase acquired in the time interval $[0,T]$ is
\begin{widetext}
\begin{eqnarray}
\theta_d^-(t)&=&-\int_0^T\frac{\dot{\widetilde{\gamma}}_2(t)\sin^2[\gamma_1(t)]}{2\cos\gamma_1(t)}dt
+\int_0^T\frac{\Theta_g\delta'(t-\frac{T}{2})\sin^2[\gamma_1(t)]}{2\cos\gamma_1(t)}dt\cr\cr
&=&-\int_0^{T/2}\frac{\dot{\widetilde{\gamma}}_2(t)\sin^2[\gamma_1(t)]}{2\cos\gamma_1(t)}dt
-\int_{T/2}^T\frac{\dot{\widetilde{\gamma}}_2(t)\sin^2[\gamma_1(t)]}{2\cos\gamma_1(t)}dt\cr\cr
&=&-\int_0^{T/2}\frac{\dot{\widetilde{\gamma}}_2(t)\sin^2[\gamma_1(t)]}{2\cos\gamma_1(t)}dt
-\int_{T/2}^0\frac{\dot{\widetilde{\gamma}}_2(\widetilde{t})\sin^2[\gamma_1(\widetilde{t})]}{2\cos\gamma_1(\widetilde{t})}d\widetilde{t}\cr\cr
&=&0.
\end{eqnarray}
\end{widetext}
Here, the condition $\gamma_1(T/2)=\pi$ is used. Besides, the geometric phase acquired in the time interval $[0,T]$ is
\begin{widetext}
\begin{eqnarray}
\theta_g^-(t)&=&-\int_0^T\dot{\gamma}_2\sin^2[\gamma_1(t)]dt\cr\cr
&=&\int_0^{T}\Theta_g\delta'(t-\frac{T}{2})\sin^2\frac{\gamma_1(t)}{2}dt-\int_0^{T}\dot{\widetilde{\gamma}}_2(t)\sin^2\frac{\gamma_1(t)}{2}dt \cr\cr
&=&\Theta_g-\int_0^{T/2}\dot{\widetilde{\gamma}}_2(t)\sin^2\frac{\gamma_1(t)}{2}dt-\int_{T/2}^{T}\dot{\widetilde{\gamma}}_2(t)\sin^2\frac{\gamma_1(t)}{2}dt \cr\cr
&=&\Theta_g-\int_0^{T/2}\dot{\widetilde{\gamma}}_2(t)\sin^2\frac{\gamma_1(t)}{2}dt-\int_{T/2}^{0}\dot{\widetilde{\gamma}}_2(\widetilde{t})\sin^2\frac{\gamma_1(\widetilde{t})}{2}d\widetilde{t} \cr\cr
&=&\Theta_g.
\end{eqnarray}
\end{widetext}
\end{appendix}
\bibliographystyle{apsrev4-1}
\bibliography{reference}

@Article{PhysRevA.107.013702,
  author    = {Qi, Shi-fan and Jing, Jun},
  journal   = {Phys. Rev. A},
  title     = {Floquet generation of a magnonic NOON state},
  year      = {2023},
  month     = {Jan},
  pages     = {013702},
  volume    = {107},
  doi       = {10.1103/PhysRevA.107.013702},
  issue     = {1},
  numpages  = {11},
  publisher = {American Physical Society},
  url       = {https://link.aps.org/doi/10.1103/PhysRevA.107.013702},
}

@Article{KYHPRA111_2025,
  author    = {Kang, Yi-Hao and Su, Qi-Ping and Wang, Yu and Shen, Lijiong and Yang, Chui-Ping},
  journal   = {Phys. Rev. A},
  title     = {Investigation of the topological quantum phase transition in a superconducting system via quantum walks},
  year      = {2025},
  month     = {Jun},
  pages     = {062209},
  volume    = {111},
  doi       = {10.1103/7dy8-1g4t},
  issue     = {6},
  numpages  = {21},
  publisher = {American Physical Society},
  url       = {https://link.aps.org/doi/10.1103/7dy8-1g4t},
}

@Article{KYHSCPMA68_2025,
  author    = {Kang, Yi-Hao and Xiao, Yang and Wang, Yu and Su, Qi-Ping and Yang, Chui-Ping},
  journal   = {Sci. China Phys. Mech. Astron.},
  title     = {Dissipation-free approach for realizing non-{H}ermitian dynamics in a superconducting circuit},
  year      = {2025},
  issn      = {1869-1927},
  month     = jan,
  number    = {3},
  pages     = {230312},
  volume    = {68},
  doi       = {10.1007/s11433-024-2549-0},
  publisher = {Springer Science and Business Media LLC},
  url       = {http://dx.doi.org/10.1007/s11433-024-2549-0},
}

@Article{Ruschhaupt_2012,
  author    = {Ruschhaupt, A and Chen, Xi and Alonso, D and Muga, J G},
  journal   = {New J. Phys.},
  title     = {Optimally robust shortcuts to population inversion in two-level quantum systems},
  year      = {2012},
  month     = {sep},
  number    = {9},
  pages     = {093040},
  volume    = {14},
  doi       = {10.1088/1367-2630/14/9/093040},
  publisher = {IOP Publishing},
  url       = {https://dx.doi.org/10.1088/1367-2630/14/9/093040},
}

@article{PhysRevLett.111.050404,
  title = {Robust Quantum Control by a Single-Shot Shaped Pulse},
  author = {Daems, D. and Ruschhaupt, A. and Sugny, D. and Gu\'erin, S.},
  journal = {Phys. Rev. Lett.},
  volume = {111},
  issue = {5},
  pages = {050404},
  numpages = {5},
  year = {2013},
  month = {Jul},
  publisher = {American Physical Society},
  doi = {10.1103/PhysRevLett.111.050404},
  url = {https://link.aps.org/doi/10.1103/PhysRevLett.111.050404}
}

@Article{PhysRevA.109.022437,
  author    = {Li, Dong-Sheng and Kang, Yi-Hao and Chen, Ye-Hong and Liu, Yang and Zhang, Cheng and Wang, Yu and Song, Jie and Xia, Yan},
  journal   = {Phys. Rev. A},
  title     = {{One-step parity measurement of $N$ cat-state qubits via reverse engineering and optimal control}},
  year      = {2024},
  month     = {Feb},
  pages     = {022437},
  volume    = {109},
  doi       = {10.1103/PhysRevA.109.022437},
  issue     = {2},
  numpages  = {12},
  publisher = {American Physical Society},
  url       = {https://link.aps.org/doi/10.1103/PhysRevA.109.022437},
}

@article{LewisJMP10_1969,
author = {Lewis, H. R. and Riesenfeld, W. B.},
title = {An Exact Quantum Theory of the Time-Dependent Harmonic Oscillator and of a Charged Particle in a Time-Dependent Electromagnetic Field},
journal = {J. Math. Phys.},
volume = {10},
number = {8},
pages = {1458-1473},
month = {Apr},
year = {1969},
doi = {10.1063/1.1664991},
URL = {https://doi.org/10.1063/1.1664991},
}

@Article{10.1063/5.0134394,
  author   = {Liu, Tong and Zhou, Yan-Hui and Wu, Qi-Cheng and Yang, Chui-Ping},
  journal  = {Appl. Phys. Lett.},
  title    = {{Generation of a GHZ-type optical entangled coherent state without measurements}},
  year     = {2022},
  issn     = {0003-6951},
  month    = {12},
  number   = {24},
  pages    = {244001},
  volume   = {121},
  doi      = {10.1063/5.0134394},
  url      = {https://doi.org/10.1063/5.0134394},
}

@Article{PhysRevA.105.042434,
  author    = {Su, Qi-Ping and Zhang, Yu and Bin, Liang and Yang, Chui-Ping},
  journal   = {Phys. Rev. A},
  title     = {Hybrid controlled-sum gate with one superconducting qutrit and one cat-state qutrit and application in hybrid entangled state preparation},
  year      = {2022},
  month     = {Apr},
  pages     = {042434},
  volume    = {105},
  doi       = {10.1103/PhysRevA.105.042434},
  issue     = {4},
  numpages  = {9},
  publisher = {American Physical Society},
  url       = {https://link.aps.org/doi/10.1103/PhysRevA.105.042434},
}

@Article{science.aaf2941,
  author   = {Chen Wang and Yvonne Y. Gao and Philip Reinhold and R. W. Heeres and Nissim Ofek and Kevin Chou and Christopher Axline and Matthew Reagor and Jacob Blumoff and K. M. Sliwa and L. Frunzio and S. M. Girvin and Liang Jiang and M. Mirrahimi and M. H. Devoret and R. J. Schoelkopf},
  journal  = {Science},
  title    = {{A Schrödinger cat living in two boxes}},
  year     = {2016},
  number   = {6289},
  pages    = {1087-1091},
  volume   = {352},
  doi      = {10.1126/science.aaf2941},
}

@article{PhysRevB.94.014506,
  title = {Quantum memory with millisecond coherence in circuit QED},
  author = {Reagor, Matthew and Pfaff, Wolfgang and Axline, Christopher and Heeres, Reinier W. and Ofek, Nissim and Sliwa, Katrina and Holland, Eric and Wang, Chen and Blumoff, Jacob and Chou, Kevin and Hatridge, Michael J. and Frunzio, Luigi and Devoret, Michel H. and Jiang, Liang and Schoelkopf, Robert J.},
  journal = {Phys. Rev. B},
  volume = {94},
  issue = {1},
  pages = {014506},
  numpages = {8},
  year = {2016},
  month = {Jul},
  publisher = {American Physical Society},
  doi = {10.1103/PhysRevB.94.014506},
  url = {https://link.aps.org/doi/10.1103/PhysRevB.94.014506}
}

@article{PRXQuantum.4.030336,
  title = {Superconducting Cavity Qubit with Tens of Milliseconds Single-Photon Coherence Time},
  author = {Milul, Ofir and Guttel, Barkay and Goldblatt, Uri and Hazanov, Sergey and Joshi, Lalit M. and Chausovsky, Daniel and Kahn, Nitzan and \ifmmode \mbox{\c{C}}\else \c{C}\fi{}ifty\"urek, Engin and Lafont, Fabien and Rosenblum, Serge},
  journal = {PRX Quantum},
  volume = {4},
  issue = {3},
  pages = {030336},
  numpages = {16},
  year = {2023},
  month = {Sep},
  publisher = {American Physical Society},
  doi = {10.1103/PRXQuantum.4.030336},
  url = {https://link.aps.org/doi/10.1103/PRXQuantum.4.030336}
}

@Article{Yan2016,
  author   = {Fei Yan and Simon Gustavsson and Archana Kamal and Jeffrey Birenbaum and Adam P. Sears and David Hover and Ted J. Gudmundsen and Danna Rosenberg and Gabriel Samach and S. Weber and Jonilyn L. Yoder and Terry P. Orlando and John Clarke and Andrew J. Kerman and William D. Oliver},
  journal  = {Nat. Commun.},
  title    = {The flux qubit revisited to enhance coherence and reproducibility},
  year     = {2016},
  month    = {nov},
  number   = {1},
  pages    = {12964},
  volume   = {7},
  doi      = {10.1038/ncomms12964},
}

@Article{5136262,
  author   = {Abdurakhimov, Leonid V. and Mahboob, Imran and Toida, Hiraku and Kakuyanagi, Kousuke and Saito, Shiro},
  journal  = {Appl. Phys. Lett.},
  title    = {{A long-lived capacitively shunted flux qubit embedded in a 3D cavity}},
  year     = {2019},
  issn     = {0003-6951},
  month    = {12},
  number   = {26},
  pages    = {262601},
  volume   = {115},
  doi      = {10.1063/1.5136262},
  url      = {https://doi.org/10.1063/1.5136262},
}

@article{PhysRevB.86.140508,
  title = {Large dispersive shift of cavity resonance induced by a superconducting flux qubit in the straddling regime},
  author = {Inomata, K. and Yamamoto, T. and Billangeon, P.-M. and Nakamura, Y. and Tsai, J. S.},
  journal = {Phys. Rev. B},
  volume = {86},
  issue = {14},
  pages = {140508},
  numpages = {4},
  year = {2012},
  month = {Oct},
  publisher = {American Physical Society},
  doi = {10.1103/PhysRevB.86.140508},
  url = {https://link.aps.org/doi/10.1103/PhysRevB.86.140508}
}

@article{PhysRevLett.115.223603,
  title = {Correlated Emission Lasing in Harmonic Oscillators Coupled via a Single Three-Level Artificial Atom},
  author = {Peng, Z. H. and Liu, Yu-xi and Peltonen, J. T. and Yamamoto, T. and Tsai, J. S. and Astafiev, O.},
  journal = {Phys. Rev. Lett.},
  volume = {115},
  issue = {22},
  pages = {223603},
  numpages = {5},
  year = {2015},
  month = {Nov},
  publisher = {American Physical Society},
  doi = {10.1103/PhysRevLett.115.223603},
  url = {https://link.aps.org/doi/10.1103/PhysRevLett.115.223603}
}

@Article{PhysRevA.109.062610,
  author    = {Xiao, Yang and Kang, Yi-Hao and Zheng, Ri-Hua and Song, Jie and Chen, Ye-Hong and Xia, Yan},
  journal   = {Phys. Rev. A},
  title     = {{Effective nonadiabatic holonomic swap gate with Rydberg atoms using invariant-based reverse engineering}},
  year      = {2024},
  month     = {Jun},
  pages     = {062610},
  volume    = {109},
  doi       = {10.1103/PhysRevA.109.062610},
  issue     = {6},
  numpages  = {12},
  publisher = {American Physical Society},
  url       = {https://link.aps.org/doi/10.1103/PhysRevA.109.062610},
}

@Article{PhysRevResearch.4.013233,
  author    = {Kang, Yi-Hao and Chen, Ye-Hong and Wang, Xin and Song, Jie and Xia, Yan and Miranowicz, Adam and Zheng, Shi-Biao and Nori, Franco},
  journal   = {Phys. Rev. Res.},
  title     = {Nonadiabatic geometric quantum computation with cat-state qubits via invariant-based reverse engineering},
  year      = {2022},
  month     = {Mar},
  pages     = {013233},
  volume    = {4},
  doi       = {10.1103/PhysRevResearch.4.013233},
  issue     = {1},
  numpages  = {16},
  publisher = {American Physical Society},
  url       = {https://link.aps.org/doi/10.1103/PhysRevResearch.4.013233},
}

@article{PhysRevA.109.042615,
  title = {Nonadiabatic geometric quantum gates by composite pulses based on superconducting qubits},
  author = {Fang, Zi-Yu and Xu, Hai and Chen, Tao and Wei, Kejin and Zhang, Chengxian},
  journal = {Phys. Rev. A},
  volume = {109},
  issue = {4},
  pages = {042615},
  numpages = {11},
  year = {2024},
  month = {Apr},
  publisher = {American Physical Society},
  doi = {10.1103/PhysRevA.109.042615},
  url = {https://link.aps.org/doi/10.1103/PhysRevA.109.042615}
}

@Article{Zhao2021,
  author   = {Zhao, PeiZi and Dong, ZhangJingZi and Zhang, ZhenXing and Guo, GuoPing and Tong, DianMin and Yin, Yi},
  journal  = {Sci. China Phys. Mech. Astron.},
  title    = {Experimental realization of nonadiabatic geometric gates with a superconducting {X}mon qubit},
  year     = {2021},
  issn     = {1869-1927},
  number   = {5},
  pages    = {250362},
  volume   = {64},
  date     = {2021-02-09},
  doi      = {10.1007/s11433-020-1641-1},
  url      = {https://doi.org/10.1007/s11433-020-1641-1},
}

@article{Kang22,
author = {Yi-Hao Kang and Jie Song and Yan Xia},
journal = {Opt. Lett.},
number = {16},
pages = {4099--4102},
publisher = {Optica Publishing Group},
title = {Error-resistant nonadiabatic binomial-code geometric quantum computation using reverse engineering},
volume = {47},
month = {Aug},
year = {2022},
url = {https://opg.optica.org/ol/abstract.cfm?URI=ol-47-16-4099},
doi = {10.1364/OL.469968},
}

@article{PhysRevResearch.3.033275,
  title = {Fast binomial-code holonomic quantum computation with ultrastrong light-matter coupling},
  author = {Chen, Ye-Hong and Qin, Wei and Stassi, Roberto and Wang, Xin and Nori, Franco},
  journal = {Phys. Rev. Res.},
  volume = {3},
  issue = {3},
  pages = {033275},
  numpages = {13},
  year = {2021},
  month = {Sep},
  publisher = {American Physical Society},
  doi = {10.1103/PhysRevResearch.3.033275},
  url = {https://link.aps.org/doi/10.1103/PhysRevResearch.3.033275}
}

@article{RevModPhys.68.733,
  title = {Quantum computation and Shor's factoring algorithm},
  author = {Ekert, Artur and Jozsa, Richard},
  journal = {Rev. Mod. Phys.},
  volume = {68},
  issue = {3},
  pages = {733--753},
  numpages = {0},
  year = {1996},
  month = {Jul},
  publisher = {American Physical Society},
  doi = {10.1103/RevModPhys.68.733},
  url = {https://link.aps.org/doi/10.1103/RevModPhys.68.733}
}

@article{PhysRevLett.79.325,
  title = {Quantum Mechanics Helps in Searching for a Needle in a Haystack},
  author = {Grover, Lov K.},
  journal = {Phys. Rev. Lett.},
  volume = {79},
  issue = {2},
  pages = {325--328},
  numpages = {0},
  year = {1997},
  month = {Jul},
  publisher = {American Physical Society},
  doi = {10.1103/PhysRevLett.79.325},
  url = {https://link.aps.org/doi/10.1103/PhysRevLett.79.325}
}

@Article{S0036144598347011,
  author   = {Shor, Peter W.},
  journal  = {SIAM Rev.},
  title    = {Polynomial-Time Algorithms for Prime Factorization and Discrete Logarithms on a Quantum Computer},
  year     = {1999},
  number   = {2},
  pages    = {303-332},
  volume   = {41},
  doi      = {10.1137/S0036144598347011},
  url      = {https://doi.org/10.1137/S0036144598347011},
}

@article{Jones2000,
  author    = {Jonathan A. Jones and Vlatko Vedral and Artur Ekert and Giuseppe Castagnoli},
  title     = {Geometric quantum computation using nuclear magnetic resonance},
  journal   = {Nature},
  year      = {2000},
  volume    = {403},
  number    = {6772},
  pages     = {869-871},
  month     = {feb},
  doi       = {10.1038/35002528},
  issn      = {1476-4687},
  url       = {https://doi.org/10.1038/35002528}
}

@article{PhysRevA.93.012306,
  title = {One-step implementation of the {R}ydberg-{R}ydberg-interaction gate},
  author = {Su, Shi-Lei and Liang, Erjun and Zhang, Shou and Wen, Jing-Ji and Sun, Li-Li and Jin, Zhao and Zhu, Ai-Dong},
  journal = {Phys. Rev. A},
  volume = {93},
  issue = {1},
  pages = {012306},
  numpages = {8},
  year = {2016},
  month = {Jan},
  publisher = {American Physical Society},
  doi = {10.1103/PhysRevA.93.012306},
  url = {https://link.aps.org/doi/10.1103/PhysRevA.93.012306}
}

@article{PhysRevA.95.022319,
  title = {{Fast Rydberg antiblockade regime and its applications in quantum logic gates}},
  author = {Su, Shi-Lei and Gao, Ya and Liang, Erjun and Zhang, Shou},
  journal = {Phys. Rev. A},
  volume = {95},
  issue = {2},
  pages = {022319},
  numpages = {11},
  year = {2017},
  month = {Feb},
  publisher = {American Physical Society},
  doi = {10.1103/PhysRevA.95.022319},
  url = {https://link.aps.org/doi/10.1103/PhysRevA.95.022319}
}

@Article{Sjqvist_2012,
  author    = {Sjöqvist, Erik and Tong, D M and Mauritz Andersson, L and Hessmo, Björn and Johansson, Markus and Singh, Kuldip},
  journal   = {New J. Phys.},
  title     = {Non-adiabatic holonomic quantum computation},
  year      = {2012},
  month     = {oct},
  number    = {10},
  pages     = {103035},
  volume    = {14},
  doi       = {10.1088/1367-2630/14/10/103035},
  publisher = {IOP Publishing},
  url       = {https://doi.org/10.1088/1367-2630/14/10/103035},
}

@article{PhysRevLett.109.170501,
  title = {Nonadiabatic Holonomic Quantum Computation in Decoherence-Free Subspaces},
  author = {Xu, G. F. and Zhang, J. and Tong, D. M. and Sj\"oqvist, Erik and Kwek, L. C.},
  journal = {Phys. Rev. Lett.},
  volume = {109},
  issue = {17},
  pages = {170501},
  numpages = {5},
  year = {2012},
  month = {Oct},
  publisher = {American Physical Society},
  doi = {10.1103/PhysRevLett.109.170501},
  url = {https://link.aps.org/doi/10.1103/PhysRevLett.109.170501}
}

@article{10.1098/rspa.1984.0023,
    author = {Berry, Michael Victor},
    title = {Quantal phase factors accompanying adiabatic changes},
    journal = {Proc. R. Soc. Lond. A.},
    volume = {392},
    number = {1802},
    pages = {45-57},
    year = {1984},
    month = {03},
    issn = {0080-4630},
    doi = {10.1098/rspa.1984.0023},
    url = {https://doi.org/10.1098/rspa.1984.0023},
   
}

@article{PhysRevA.92.022320,
  title = {Universal holonomic quantum gates in decoherence-free subspace on superconducting circuits},
  author = {Xue, Zheng-Yuan and Zhou, Jian and Wang, Z. D.},
  journal = {Phys. Rev. A},
  volume = {92},
  issue = {2},
  pages = {022320},
  numpages = {6},
  year = {2015},
  month = {Aug},
  publisher = {American Physical Society},
  doi = {10.1103/PhysRevA.92.022320},
  url = {https://link.aps.org/doi/10.1103/PhysRevA.92.022320}
}

@article{PhysRevLett.58.1593,
  title = {Phase change during a cyclic quantum evolution},
  author = {Aharonov, Y. and Anandan, J.},
  journal = {Phys. Rev. Lett.},
  volume = {58},
  issue = {16},
  pages = {1593--1596},
  numpages = {0},
  year = {1987},
  month = {Apr},
  publisher = {American Physical Society},
  doi = {10.1103/PhysRevLett.58.1593},
  url = {https://link.aps.org/doi/10.1103/PhysRevLett.58.1593}
}

@article{PhysRevA.95.032311,
  title = {Composite nonadiabatic holonomic quantum computation},
  author = {Xu, G. F. and Zhao, P. Z. and Xing, T. H. and Sj\"oqvist, Erik and Tong, D. M.},
  journal = {Phys. Rev. A},
  volume = {95},
  issue = {3},
  pages = {032311},
  numpages = {6},
  year = {2017},
  month = {Mar},
  publisher = {American Physical Society},
  doi = {10.1103/PhysRevA.95.032311},
  url = {https://link.aps.org/doi/10.1103/PhysRevA.95.032311}
}

@article{PhysRevA.101.022330,
  title = {Noncyclic geometric quantum computation with shortcut to adiabaticity},
  author = {Lv, Qing-Xian and Liang, Zhen-Tao and Liu, Hong-Zhi and Liang, Jia-Hao and Liao, Kai-Yu and Du, Yan-Xiong},
  journal = {Phys. Rev. A},
  volume = {101},
  issue = {2},
  pages = {022330},
  numpages = {7},
  year = {2020},
  month = {Feb},
  publisher = {American Physical Society},
  doi = {10.1103/PhysRevA.101.022330},
  url = {https://link.aps.org/doi/10.1103/PhysRevA.101.022330}
}

@article{PhysRevA.95.062310,
  title = {Single-shot realization of nonadiabatic holonomic quantum gates in decoherence-free subspaces},
  author = {Zhao, P. Z. and Xu, G. F. and Ding, Q. M. and Sj\"oqvist, Erik and Tong, D. M.},
  journal = {Phys. Rev. A},
  volume = {95},
  issue = {6},
  pages = {062310},
  numpages = {6},
  year = {2017},
  month = {Jun},
  publisher = {American Physical Society},
  doi = {10.1103/PhysRevA.95.062310},
  url = {https://link.aps.org/doi/10.1103/PhysRevA.95.062310}
}

@article{PhysRevA.98.052315,
  title = {Path-shortening realizations of nonadiabatic holonomic gates},
  author = {Xu, G. F. and Tong, D. M. and Sj\"oqvist, Erik},
  journal = {Phys. Rev. A},
  volume = {98},
  issue = {5},
  pages = {052315},
  numpages = {5},
  year = {2018},
  month = {Nov},
  publisher = {American Physical Society},
  doi = {10.1103/PhysRevA.98.052315},
  url = {https://link.aps.org/doi/10.1103/PhysRevA.98.052315}
}

@article{PhysRevLett.123.100501,
  title = {Plug-and-Play Approach to Nonadiabatic Geometric Quantum Gates},
  author = {Liu, Bao-Jie and Song, Xue-Ke and Xue, Zheng-Yuan and Wang, Xin and Yung, Man-Hong},
  journal = {Phys. Rev. Lett.},
  volume = {123},
  issue = {10},
  pages = {100501},
  numpages = {6},
  year = {2019},
  month = {Sep},
  publisher = {American Physical Society},
  doi = {10.1103/PhysRevLett.123.100501},
  url = {https://link.aps.org/doi/10.1103/PhysRevLett.123.100501}
}

@article{PhysRevA.100.012329,
  title = {Searching nonadiabatic holonomic quantum gates via an optimization algorithm},
  author = {Zhang, Feihao and Zhang, Jiang and Gao, Pan and Long, Guilu},
  journal = {Phys. Rev. A},
  volume = {100},
  issue = {1},
  pages = {012329},
  numpages = {12},
  year = {2019},
  month = {Jul},
  publisher = {American Physical Society},
  doi = {10.1103/PhysRevA.100.012329},
  url = {https://link.aps.org/doi/10.1103/PhysRevA.100.012329}
}

@article{PhysRevA.101.032322,
  title = {Flexible scheme for the implementation of nonadiabatic geometric quantum computation},
  author = {Kang, Yi-Hao and Shi, Zhi-Cheng and Huang, Bi-Hua and Song, Jie and Xia, Yan},
  journal = {Phys. Rev. A},
  volume = {101},
  issue = {3},
  pages = {032322},
  numpages = {16},
  year = {2020},
  month = {Mar},
  publisher = {American Physical Society},
  doi = {10.1103/PhysRevA.101.032322},
  url = {https://link.aps.org/doi/10.1103/PhysRevA.101.032322}
}

@article{PhysRevA.52.R2493,
  title = {Scheme for reducing decoherence in quantum computer memory},
  author = {Shor, Peter W.},
  journal = {Phys. Rev. A},
  volume = {52},
  issue = {4},
  pages = {R2493--R2496},
  numpages = {0},
  year = {1995},
  month = {Oct},
  publisher = {American Physical Society},
  doi = {10.1103/PhysRevA.52.R2493},
  url = {https://link.aps.org/doi/10.1103/PhysRevA.52.R2493}
}

@article{PhysRevA.86.032324,
  title = {Surface codes: Towards practical large-scale quantum computation},
  author = {Fowler, Austin G. and Mariantoni, Matteo and Martinis, John M. and Cleland, Andrew N.},
  journal = {Phys. Rev. A},
  volume = {86},
  issue = {3},
  pages = {032324},
  numpages = {48},
  year = {2012},
  month = {Sep},
  publisher = {American Physical Society},
  doi = {10.1103/PhysRevA.86.032324},
  url = {https://link.aps.org/doi/10.1103/PhysRevA.86.032324}
}

@article{PhysRevLett.119.180509,
  title = {Error Mitigation for Short-Depth Quantum Circuits},
  author = {Temme, Kristan and Bravyi, Sergey and Gambetta, Jay M.},
  journal = {Phys. Rev. Lett.},
  volume = {119},
  issue = {18},
  pages = {180509},
  numpages = {5},
  year = {2017},
  month = {Nov},
  publisher = {American Physical Society},
  doi = {10.1103/PhysRevLett.119.180509},
  url = {https://link.aps.org/doi/10.1103/PhysRevLett.119.180509}
}

@article{Kandala2019,
  author    = {Abhinav Kandala and Kristan Temme and Antonio D. C{\'o}rcoles and Antonio Mezzacapo and Jerry M. Chow and Jay M. Gambetta},
  title     = {Error mitigation extends the computational reach of a noisy quantum processor},
  journal   = {Nature},
  year      = {2019},
  volume    = {567},
  number    = {7749},
  pages     = {491--495},
  month     = {mar},
  doi       = {10.1038/s41586-019-1040-7},
  issn      = {1476-4687},
  url       = {https://www.nature.com/articles/s41586-019-1040-7}
}

@Article{Peruzzo2014,
  author   = {Alberto Peruzzo and Jarrod McClean and Peter Shadbolt and Man-Hong Yung and Xiao-Qi Zhou and Peter J. Love and Al{\'a}n Aspuru-Guzik and Jeremy L. O'Brien},
  journal  = {Nat. Commun.},
  title    = {A variational eigenvalue solver on a photonic quantum processor},
  year     = {2014},
  issn     = {2041-1723},
  month    = {jul},
  number   = {1},
  pages    = {4213},
  volume   = {5},
  doi      = {10.1038/ncomms5213},
  url      = {https://doi.org/10.1038/ncomms5213},
}

@article{Kandala2017,
  author    = {Abhinav Kandala and Antonio Mezzacapo and Kristan Temme and Maika Takita and Markus Brink and Jerry M. Chow and Jay M. Gambetta},
  title     = {Hardware-efficient variational quantum eigensolver for small molecules and quantum magnets},
  journal   = {Nature},
  year      = {2017},
  volume    = {549},
  number    = {7671},
  pages     = {242--246},
  month     = {sep},
  doi       = {10.1038/nature23879},
  issn      = {1476-4687},
  url       = {https://www.nature.com/articles/nature23879}
}

@article{PhysRevLett.125.250403,
  title = {Optimal Robust Quantum Control by Inverse Geometric Optimization},
  author = {Dridi, Ghassen and Liu, Kaipeng and Gu\'erin, St\'ephane},
  journal = {Phys. Rev. Lett.},
  volume = {125},
  issue = {25},
  pages = {250403},
  numpages = {6},
  year = {2020},
  month = {Dec},
  publisher = {American Physical Society},
  doi = {10.1103/PhysRevLett.125.250403},
  url = {https://link.aps.org/doi/10.1103/PhysRevLett.125.250403}
}

@Article{doi:10.1126/sciadv.adr0875,
  author   = {Zi-Jie Chen and Hongwei Huang and Lida Sun and Qing-Xuan Jie and Jie Zhou and Ziyue Hua and Yifang Xu and Weiting Wang and Guang-Can Guo and Chang-Ling Zou and Luyan Sun and Xu-Bo Zou},
  journal  = {Sci. Adv.},
  title    = {Robust and optimal control of open quantum systems},
  year     = {2025},
  number   = {9},
  pages    = {eadr0875},
  volume   = {11},
  doi      = {10.1126/sciadv.adr0875},
  url      = {https://www.science.org/doi/abs/10.1126/sciadv.adr0875},
}

@Article{Heeres2017,
  author   = {Reinier W. Heeres and Philip Reinhold and Nissim Ofek and Luigi Frunzio and Liang Jiang and Michel H. Devoret and Robert J. Schoelkopf},
  journal  = {Nat. Commun.},
  title    = {Implementing a universal gate set on a logical qubit encoded in an oscillator},
  year     = {2017},
  issn     = {2041-1723},
  month    = {jul},
  number   = {1},
  pages    = {94},
  volume   = {8},
  doi      = {10.1038/s41467-017-00045-1},
  url      = {https://www.nature.com/articles/s41467-017-00045-1},
}

@article{CAI202150,
title = {Bosonic quantum error correction codes in superconducting quantum circuits},
journal = {Fundam. Res.},
volume = {1},
number = {1},
pages = {50-67},
year = {2021},
issn = {2667-3258},
doi = {https://doi.org/10.1016/j.fmre.2020.12.006},
url = {https://www.sciencedirect.com/science/article/pii/S2667325820300145},
author = {Weizhou Cai and Yuwei Ma and Weiting Wang and Chang-Ling Zou and Luyan Sun}
}

@article{vf7v-cpq9,
  title = {Fault-Tolerant Logical {C}lifford Gates from Code Automorphisms},
  author = {Sayginel, Hasan and Koutsioumpas, Stergios and Webster, Mark and Rajput, Abhishek and Browne, Dan E.},
  journal = {PRX Quantum},
  volume = {6},
  issue = {3},
  pages = {030343},
  numpages = {24},
  year = {2025},
  month = {Sep},
  publisher = {American Physical Society},
  doi = {10.1103/vf7v-cpq9},
  url = {https://link.aps.org/doi/10.1103/vf7v-cpq9}
}

@article{PhysRevX.10.011058,
  title = {Quantum Computing with Rotation-Symmetric Bosonic Codes},
  author = {Grimsmo, Arne L. and Combes, Joshua and Baragiola, Ben Q.},
  journal = {Phys. Rev. X},
  volume = {10},
  issue = {1},
  pages = {011058},
  numpages = {32},
  year = {2020},
  month = {Mar},
  publisher = {American Physical Society},
  doi = {10.1103/PhysRevX.10.011058},
  url = {https://link.aps.org/doi/10.1103/PhysRevX.10.011058}
}

@article{PhysRevA.97.032346,
  title = {Performance and structure of single-mode bosonic codes},
  author = {Albert, Victor V. and Noh, Kyungjoo and Duivenvoorden, Kasper and Young, Dylan J. and Brierley, R. T. and Reinhold, Philip and Vuillot, Christophe and Li, Linshu and Shen, Chao and Girvin, S. M. and Terhal, Barbara M. and Jiang, Liang},
  journal = {Phys. Rev. A},
  volume = {97},
  issue = {3},
  pages = {032346},
  numpages = {30},
  year = {2018},
  month = {Mar},
  publisher = {American Physical Society},
  doi = {10.1103/PhysRevA.97.032346},
  url = {https://link.aps.org/doi/10.1103/PhysRevA.97.032346}
}

@Article{doi:10.1126/science.1243289,
  author   = {Brian Vlastakis and Gerhard Kirchmair and Zaki Leghtas and Simon E. Nigg and Luigi Frunzio and S. M. Girvin and Mazyar Mirrahimi and M. H. Devoret and R. J. Schoelkopf},
  journal  = {Science},
  title    = {Deterministically Encoding Quantum Information Using 100-Photon {Schrödinger} Cat States},
  year     = {2013},
  number   = {6158},
  pages    = {607-610},
  volume   = {342},
  doi      = {10.1126/science.1243289},
  url      = {https://www.science.org/doi/abs/10.1126/science.1243289},
}

@article{Albert_2019,
doi = {10.1088/2058-9565/ab1e69},
url = {https://doi.org/10.1088/2058-9565/ab1e69},
year = {2019},
month = {jun},
publisher = {IOP Publishing},
volume = {4},
number = {3},
pages = {035007},
author = {Albert, Victor V and Mundhada, Shantanu O and Grimm, Alexander and Touzard, Steven and Devoret, Michel H and Jiang, Liang},
title = {Pair-cat codes: autonomous error-correction with low-order nonlinearity},
journal = {Quantum Sci. Technol.}
}

@article{PhysRevX.6.031006,
  title = {New Class of Quantum Error-Correcting Codes for a Bosonic Mode},
  author = {Michael, Marios H. and Silveri, Matti and Brierley, R. T. and Albert, Victor V. and Salmilehto, Juha and Jiang, Liang and Girvin, S. M.},
  journal = {Phys. Rev. X},
  volume = {6},
  issue = {3},
  pages = {031006},
  numpages = {26},
  year = {2016},
  month = {Jul},
  publisher = {American Physical Society},
  doi = {10.1103/PhysRevX.6.031006},
  url = {https://link.aps.org/doi/10.1103/PhysRevX.6.031006}
}

@article{PhysRevA.54.3824,
  title = {Mixed-state entanglement and quantum error correction},
  author = {Bennett, Charles H. and DiVincenzo, David P. and Smolin, John A. and Wootters, William K.},
  journal = {Phys. Rev. A},
  volume = {54},
  issue = {5},
  pages = {3824--3851},
  numpages = {0},
  year = {1996},
  month = {Nov},
  publisher = {American Physical Society},
  doi = {10.1103/PhysRevA.54.3824},
  url = {https://link.aps.org/doi/10.1103/PhysRevA.54.3824}
}

@article{PhysRevA.55.900,
  title = {Theory of quantum error-correcting codes},
  author = {Knill, Emanuel and Laflamme, Raymond},
  journal = {Phys. Rev. A},
  volume = {55},
  issue = {2},
  pages = {900--911},
  numpages = {0},
  year = {1997},
  month = {Feb},
  publisher = {American Physical Society},
  doi = {10.1103/PhysRevA.55.900},
  url = {https://link.aps.org/doi/10.1103/PhysRevA.55.900}
}

@article{PhysRevA.62.062303,
  title = {Family of {G}rover's quantum-searching algorithms},
  author = {Galindo, A. and Mart\'{\i}n-Delgado, M. A.},
  journal = {Phys. Rev. A},
  volume = {62},
  issue = {6},
  pages = {062303},
  numpages = {6},
  year = {2000},
  month = {Nov},
  publisher = {American Physical Society},
  doi = {10.1103/PhysRevA.62.062303},
  url = {https://link.aps.org/doi/10.1103/PhysRevA.62.062303}
}

@article{PhysRevA.103.012205,
  title = {Dynamical-decoupling-protected nonadiabatic holonomic quantum computation},
  author = {Zhao, P. Z. and Wu, X. and Tong, D. M.},
  journal = {Phys. Rev. A},
  volume = {103},
  issue = {1},
  pages = {012205},
  numpages = {6},
  year = {2021},
  month = {Jan},
  publisher = {American Physical Society},
  doi = {10.1103/PhysRevA.103.012205},
  url = {https://link.aps.org/doi/10.1103/PhysRevA.103.012205}
}

@article{Zhou15,
author = {Jian Zhou and Wei-Can Yu and Yu-Mei Gao and Zheng-Yuan Xue},
journal = {Opt. Express},
number = {11},
pages = {14027--14035},
publisher = {Optica Publishing Group},
title = {Cavity QED implementation of non-adiabatic holonomies for universal quantum gates in decoherence-free subspaces with nitrogen-vacancy centers},
volume = {23},
month = {Jun},
year = {2015},
url = {https://opg.optica.org/oe/abstract.cfm?URI=oe-23-11-14027},
doi = {10.1364/OE.23.014027},
}

@article{PhysRevA.97.022335,
  title = {Holonomic surface codes for fault-tolerant quantum computation},
  author = {Zhang, Jiang and Devitt, Simon J. and You, J. Q. and Nori, Franco},
  journal = {Phys. Rev. A},
  volume = {97},
  issue = {2},
  pages = {022335},
  numpages = {13},
  year = {2018},
  month = {Feb},
  publisher = {American Physical Society},
  doi = {10.1103/PhysRevA.97.022335},
  url = {https://link.aps.org/doi/10.1103/PhysRevA.97.022335}
}

@article{10.1143/PTP.70.439,
    author = {Suzuki, Kenji and Okamoto, Ryoji},
    title = {Degenerate perturbation theory in quantum mechanics},
    journal = {Prog. Theor. Phys.},
    volume = {70},
    number = {2},
    pages = {439-451},
    year = {1983},
    month = {08},
    issn = {0033-068X},
    doi = {10.1143/PTP.70.439},
    url = {https://doi.org/10.1143/PTP.70.439}
}

@article{PhysRevLett.124.120501,
  title = {Demonstration of Controlled-Phase Gates between Two Error-Correctable Photonic Qubits},
  author = {Xu, Y. and Ma, Y. and Cai, W. and Mu, X. and Dai, W. and Wang, W. and Hu, L. and Li, X. and Han, J. and Wang, H. and Song, Y. P. and Yang, Zhen-Biao and Zheng, Shi-Biao and Sun, L.},
  journal = {Phys. Rev. Lett.},
  volume = {124},
  issue = {12},
  pages = {120501},
  numpages = {7},
  year = {2020},
  month = {Mar},
  publisher = {American Physical Society},
  doi = {10.1103/PhysRevLett.124.120501},
  url = {https://link.aps.org/doi/10.1103/PhysRevLett.124.120501}
}

@Article{Hu2019,
  author  = {Hu, L. and Ma, Y. and Cai, W. and Mu, X. and Xu, Y. and Wang, W. and Wu, Y. and Wang, H. and Song, Y. P. and Zou, C.-L. and Girvin, S. M. and Duan, L.-M. and Sun, L.},
  journal = {Nat. Phys.},
  title   = {Quantum error correction and universal gate set operation on a binomial bosonic logical qubit},
  year    = {2019},
  issn    = {1745-2481},
  month   = {5},
  number  = {5},
  pages   = {503--508},
  volume  = {15},
  date    = {2019/05/01},
  doi     = {10.1038/s41567-018-0414-3},
  url     = {https://doi.org/10.1038/s41567-018-0414-3},
}

@Article{Chou2018,
  author  = {Chou, Kevin S. and Blumoff, Jacob Z. and Wang, Christopher S. and Reinhold, Philip C. and Axline, Christopher J. and Gao, Yvonne Y. and Frunzio, L. and Devoret, M. H. and Jiang, Liang and Schoelkopf, R. J.},
  journal = {Nature},
  title   = {Deterministic teleportation of a quantum gate between two logical qubits},
  year    = {2018},
  issn    = {1476-4687},
  month   = {9},
  number  = {7723},
  pages   = {368--373},
  volume  = {561},
  date    = {2018/09/01},
  doi     = {10.1038/s41586-018-0470-y},
  url     = {https://www.nature.com/articles/s41586-018-0470-y},
}

@Article{Gao2019,
  author  = {Gao, Yvonne Y. and Lester, Brian J. and Chou, Kevin S. and Frunzio, Luigi and Devoret, Michel H. and Jiang, Liang and Girvin, S. M. and Schoelkopf, Robert J.},
  journal = {Nature},
  title   = {Entanglement of bosonic modes through an engineered exchange interaction},
  year    = {2019},
  issn    = {1476-4687},
  month   = {2},
  number  = {7745},
  pages   = {509--512},
  volume  = {566},
  date    = {2019/02/01},
  doi     = {10.1038/s41586-019-0970-4},
  url     = {https://www.nature.com/articles/s41586-019-0970-4},
}

@Article{Ma2020,
  author  = {Ma, Y. and Xu, Y. and Mu, X. and Cai, W. and Hu, L. and Wang, W. and Pan, X. and Wang, H. and Song, Y. P. and Zou, C.-L. and Sun, L.},
  journal = {Nat. Phys.},
  title   = {Error-transparent operations on a logical qubit protected by quantum error correction},
  year    = {2020},
  issn    = {1745-2481},
  month   = {8},
  number  = {8},
  pages   = {827--831},
  volume  = {16},
  date    = {2020/08/01},
  doi     = {10.1038/s41567-020-0893-x},
  url     = {https://www.nature.com/articles/s41567-020-0893-x},
}

@Article{Vandersypen2001,
  author  = {Vandersypen, Lieven M. K. and Steffen, Matthias and Breyta, Gregory and Yannoni, Costantino S. and Sherwood, Mark H. and Chuang, Isaac L.},
  journal = {Nature},
  title   = {Experimental realization of {S}hor's quantum factoring algorithm using nuclear magnetic resonance},
  year    = {2001},
  issn    = {1476-4687},
  month   = {12},
  number  = {6866},
  pages   = {883--887},
  volume  = {414},
  date    = {2001/12/01},
  doi     = {10.1038/414883a},
  url     = {https://www.nature.com/articles/414883a},
}

@article{PhysRevLett.108.130501,
  title = {Quantum Factorization of 143 on a Dipolar-Coupling Nuclear Magnetic Resonance System},
  author = {Xu, Nanyang and Zhu, Jing and Lu, Dawei and Zhou, Xianyi and Peng, Xinhua and Du, Jiangfeng},
  journal = {Phys. Rev. Lett.},
  volume = {108},
  issue = {13},
  pages = {130501},
  numpages = {5},
  year = {2012},
  month = {Mar},
  publisher = {American Physical Society},
  doi = {10.1103/PhysRevLett.108.130501},
  url = {https://link.aps.org/doi/10.1103/PhysRevLett.108.130501}
}

@article{PhysRevA.105.062455,
  title = {Hard-instance learning for quantum adiabatic prime factorization},
  author = {Lin, Jian and Zhang, Zhengfeng and Zhang, Junping and Li, Xiaopeng},
  journal = {Phys. Rev. A},
  volume = {105},
  issue = {6},
  pages = {062455},
  numpages = {9},
  year = {2022},
  month = {Jun},
  publisher = {American Physical Society},
  doi = {10.1103/PhysRevA.105.062455},
  url = {https://link.aps.org/doi/10.1103/PhysRevA.105.062455}
}

@Article{WOS001177135800178,
  author               = {Zheng, Xiaodong and Wang, Jianhui and Yue, Meng},
  journal              = {IEEE Trans. Power Syst.},
  title                = {A Fast Quantum Algorithm for Searching the Quasi-Optimal Solutions of Unit Commitment},
  year                 = {2024},
  issn                 = {0885-8950},
  month                = {MAR},
  number               = {2},
  pages                = {4755-4758},
  volume               = {39},
  doi                  = {10.1109/TPWRS.2024.3350382},
  eissn                = {1558-0679},
  orcid-numbers        = {Zheng, Xiaodong/0000-0002-7993-9963},
  researcherid-numbers = {Zheng, Xiaodong/ISC-0047-2023 Wang, jianhui/B-5187-2013},
  unique-id            = {WOS:001177135800178},
}

@Article{Kang_2023,
  author    = {Kang, Yi-Hao and Xiao, Yang and Shi, Zhi-Cheng and Wang, Yu and Yang, Jian-Qun and Song, Jie and Xia, Yan},
  journal   = {New J. Phys.},
  title     = {Effective implementation of nonadiabatic geometric quantum gates of cat-state qubits using an auxiliary qutrit},
  year      = {2023},
  month     = {mar},
  number    = {3},
  pages     = {033029},
  volume    = {25},
  doi       = {10.1088/1367-2630/acc2de},
  publisher = {IOP Publishing},
  url       = {https://doi.org/10.1088/1367-2630/acc2de},
}

@article{PhysRevA.96.022309,
  title = {Robust not gate by single-shot-shaped pulses: Demonstration of the efficiency of the pulses in rephasing atomic coherences},
  author = {Van-Damme, Leo and Schraft, Daniel and Genov, Genko T. and Sugny, Dominique and Halfmann, Thomas and Gu\'erin, St\'ephane},
  journal = {Phys. Rev. A},
  volume = {96},
  issue = {2},
  pages = {022309},
  numpages = {8},
  year = {2017},
  month = {Aug},
  publisher = {American Physical Society},
  doi = {10.1103/PhysRevA.96.022309},
  url = {https://link.aps.org/doi/10.1103/PhysRevA.96.022309}
}

@article{PhysRevA.97.042336,
  title = {{Nonadiabatic holonomic quantum computation using Rydberg blockade}},
  author = {Kang, Yi-Hao and Chen, Ye-Hong and Shi, Zhi-Cheng and Huang, Bi-Hua and Song, Jie and Xia, Yan},
  journal = {Phys. Rev. A},
  volume = {97},
  issue = {4},
  pages = {042336},
  numpages = {12},
  year = {2018},
  month = {Apr},
  publisher = {American Physical Society},
  doi = {10.1103/PhysRevA.97.042336},
  url = {https://link.aps.org/doi/10.1103/PhysRevA.97.042336}
}

@article{PhysRevA.100.043413,
  title = {Robust and highly efficient discrimination of chiral molecules through three-mode parallel paths},
  author = {Wu, Jin-Lei and Wang, Yan and Song, Jie and Xia, Yan and Su, Shi-Lei and Jiang, Yong-Yuan},
  journal = {Phys. Rev. A},
  volume = {100},
  issue = {4},
  pages = {043413},
  numpages = {11},
  year = {2019},
  month = {Oct},
  publisher = {American Physical Society},
  doi = {10.1103/PhysRevA.100.043413},
  url = {https://link.aps.org/doi/10.1103/PhysRevA.100.043413}
}

@article{PhysRevA.92.052302,
  title = {Nonadiabatic holonomic gates realized by a single-shot implementation},
  author = {Xu, G. F. and Liu, C. L. and Zhao, P. Z. and Tong, D. M.},
  journal = {Phys. Rev. A},
  volume = {92},
  issue = {5},
  pages = {052302},
  numpages = {5},
  year = {2015},
  month = {Nov},
  publisher = {American Physical Society},
  doi = {10.1103/PhysRevA.92.052302},
  url = {https://link.aps.org/doi/10.1103/PhysRevA.92.052302}
}

@article{PhysRevApplied.10.054051,
  title = {Nonadiabatic Geometric Quantum Computation with Parametrically Tunable Coupling},
  author = {Chen, Tao and Xue, Zheng-Yuan},
  journal = {Phys. Rev. Appl.},
  volume = {10},
  issue = {5},
  pages = {054051},
  numpages = {13},
  year = {2018},
  month = {Nov},
  publisher = {American Physical Society},
  doi = {10.1103/PhysRevApplied.10.054051},
  url = {https://link.aps.org/doi/10.1103/PhysRevApplied.10.054051}
}

@article{PhysRevLett.123.200502,
  title = {All-Gaussian Universality and Fault Tolerance with the {G}ottesman-{K}itaev-{P}reskill Code},
  author = {Baragiola, Ben Q. and Pantaleoni, Giacomo and Alexander, Rafael N. and Karanjai, Angela and Menicucci, Nicolas C.},
  journal = {Phys. Rev. Lett.},
  volume = {123},
  issue = {20},
  pages = {200502},
  numpages = {6},
  year = {2019},
  month = {Nov},
  publisher = {American Physical Society},
  doi = {10.1103/PhysRevLett.123.200502},
  url = {https://link.aps.org/doi/10.1103/PhysRevLett.123.200502}
}

@article{PhysRevA.96.032341,
  title = {Fault tolerance with bare ancillary qubits for a [[7,1,3]] code},
  author = {Li, Muyuan and Guti\'errez, Mauricio and David, Stanley E. and Hernandez, Alonzo and Brown, Kenneth R.},
  journal = {Phys. Rev. A},
  volume = {96},
  issue = {3},
  pages = {032341},
  numpages = {10},
  year = {2017},
  month = {Sep},
  publisher = {American Physical Society},
  doi = {10.1103/PhysRevA.96.032341},
  url = {https://link.aps.org/doi/10.1103/PhysRevA.96.032341}
}

@article{PhysRevA.56.1114,
  title = {Bosonic quantum codes for amplitude damping},
  author = {Chuang, Isaac L. and Leung, Debbie W. and Yamamoto, Yoshihisa},
  journal = {Phys. Rev. A},
  volume = {56},
  issue = {2},
  pages = {1114--1125},
  numpages = {0},
  year = {1997},
  month = {Aug},
  publisher = {American Physical Society},
  doi = {10.1103/PhysRevA.56.1114},
  url = {https://link.aps.org/doi/10.1103/PhysRevA.56.1114}
}

@article{PhysRevA.64.012310,
  title = {Encoding a qubit in an oscillator},
  author = {Gottesman, Daniel and Kitaev, Alexei and Preskill, John},
  journal = {Phys. Rev. A},
  volume = {64},
  issue = {1},
  pages = {012310},
  numpages = {21},
  year = {2001},
  month = {Jun},
  publisher = {American Physical Society},
  doi = {10.1103/PhysRevA.64.012310},
  url = {https://link.aps.org/doi/10.1103/PhysRevA.64.012310}
}

@article{PhysRevLett.116.140502,
  title = {Holonomic Quantum Control with Continuous Variable Systems},
  author = {Albert, Victor V. and Shu, Chi and Krastanov, Stefan and Shen, Chao and Liu, Ren-Bao and Yang, Zhen-Biao and Schoelkopf, Robert J. and Mirrahimi, Mazyar and Devoret, Michel H. and Jiang, Liang},
  journal = {Phys. Rev. Lett.},
  volume = {116},
  issue = {14},
  pages = {140502},
  numpages = {6},
  year = {2016},
  month = {Apr},
  publisher = {American Physical Society},
  doi = {10.1103/PhysRevLett.116.140502},
  url = {https://link.aps.org/doi/10.1103/PhysRevLett.116.140502}
}

@article{PhysRevX.9.041053,
  title = {Repetition Cat Qubits for Fault-Tolerant Quantum Computation},
  author = {Guillaud, J\'er\'emie and Mirrahimi, Mazyar},
  journal = {Phys. Rev. X},
  volume = {9},
  issue = {4},
  pages = {041053},
  numpages = {23},
  year = {2019},
  month = {Dec},
  publisher = {American Physical Society},
  doi = {10.1103/PhysRevX.9.041053},
  url = {https://link.aps.org/doi/10.1103/PhysRevX.9.041053}
}

@article{PhysRevLett.126.023602,
  title = {Shortcuts to Adiabaticity for the Quantum Rabi Model: Efficient Generation of Giant Entangled Cat States via Parametric Amplification},
  author = {Chen, Ye-Hong and Qin, Wei and Wang, Xin and Miranowicz, Adam and Nori, Franco},
  journal = {Phys. Rev. Lett.},
  volume = {126},
  issue = {2},
  pages = {023602},
  numpages = {8},
  year = {2021},
  month = {Jan},
  publisher = {American Physical Society},
  doi = {10.1103/PhysRevLett.126.023602},
  url = {https://link.aps.org/doi/10.1103/PhysRevLett.126.023602}
}

@article{PhysRevLett.119.030502,
  title = {Cat Codes with Optimal Decoherence Suppression for a Lossy Bosonic Channel},
  author = {Li, Linshu and Zou, Chang-Ling and Albert, Victor V. and Muralidharan, Sreraman and Girvin, S. M. and Jiang, Liang},
  journal = {Phys. Rev. Lett.},
  volume = {119},
  issue = {3},
  pages = {030502},
  numpages = {6},
  year = {2017},
  month = {Jul},
  publisher = {American Physical Society},
  doi = {10.1103/PhysRevLett.119.030502},
  url = {https://link.aps.org/doi/10.1103/PhysRevLett.119.030502}
}

@article{PhysRevA.84.043832,
  title = {Dissipation and ultrastrong coupling in circuit QED},
  author = {Beaudoin, F\'elix and Gambetta, Jay M. and Blais, A.},
  journal = {Phys. Rev. A},
  volume = {84},
  issue = {4},
  pages = {043832},
  numpages = {15},
  year = {2011},
  month = {Oct},
  publisher = {American Physical Society},
  doi = {10.1103/PhysRevA.84.043832},
  url = {https://link.aps.org/doi/10.1103/PhysRevA.84.043832}
}

@article{PhysRevLett.110.243601,
  title = {Spontaneous Conversion from Virtual to Real Photons in the Ultrastrong-Coupling Regime},
  author = {Stassi, R. and Ridolfo, A. and Di Stefano, O. and Hartmann, M. J. and Savasta, S.},
  journal = {Phys. Rev. Lett.},
  volume = {110},
  issue = {24},
  pages = {243601},
  numpages = {5},
  year = {2013},
  month = {Jun},
  publisher = {American Physical Society},
  doi = {10.1103/PhysRevLett.110.243601},
  url = {https://link.aps.org/doi/10.1103/PhysRevLett.110.243601}
}

@article{PhysRevA.83.062116,
  title = {{Lewis-Riesenfeld} invariants and transitionless quantum driving},
  author = {Chen, Xi and Torrontegui, E. and Muga, J. G.},
  journal = {Phys. Rev. A},
  volume = {83},
  issue = {6},
  pages = {062116},
  numpages = {8},
  year = {2011},
  month = {Jun},
  publisher = {American Physical Society},
  doi = {10.1103/PhysRevA.83.062116},
  url = {https://link.aps.org/doi/10.1103/PhysRevA.83.062116}
}

@article{PhysRevA.86.033405,
  title = {Engineering of fast population transfer in three-level systems},
  author = {Chen, Xi and Muga, J. G.},
  journal = {Phys. Rev. A},
  volume = {86},
  issue = {3},
  pages = {033405},
  numpages = {6},
  year = {2012},
  month = {Sep},
  publisher = {American Physical Society},
  doi = {10.1103/PhysRevA.86.033405},
  url = {https://link.aps.org/doi/10.1103/PhysRevA.86.033405}
}

@article{PhysRevA.89.043408,
  title = {Hamiltonian engineering via invariants and dynamical algebra},
  author = {Torrontegui, E. and Mart\'{\i}nez-Garaot, S. and Muga, J. G.},
  journal = {Phys. Rev. A},
  volume = {89},
  issue = {4},
  pages = {043408},
  numpages = {8},
  year = {2014},
  month = {Apr},
  publisher = {American Physical Society},
  doi = {10.1103/PhysRevA.89.043408},
  url = {https://link.aps.org/doi/10.1103/PhysRevA.89.043408}
}

@article{Leghtas2015,
author = {Leghtas, Z and Touzard, S and Pop, I M and Kou, A and Vlastakis, B and Petrenko, A and Sliwa, K M and Narla, A and Shankar, S and Hatridge, M J and Reagor, M and Frunzio, L and Schoelkopf, R J and Mirrahimi, M and Devoret, M H},
title = {Quantum engineering. Confining the state of light to a quantum manifold by engineered two-photon loss},
journal = {Science},
year = {2015},
volume = {347},
number = {6224},
pages = {853--857},
doi = {10.1126/science.aaa2085},
month = {feb}
}

@article{PhysRevA.97.012333,
  title = {Quantum state transfer in spin chains via shortcuts to adiabaticity},
  author = {Huang, Bi-Hua and Kang, Yi-Hao and Chen, Ye-Hong and Shi, Zhi-Cheng and Song, Jie and Xia, Yan},
  journal = {Phys. Rev. A},
  volume = {97},
  issue = {1},
  pages = {012333},
  numpages = {9},
  year = {2018},
  month = {Jan},
  publisher = {American Physical Society},
  doi = {10.1103/PhysRevA.97.012333},
  url = {https://link.aps.org/doi/10.1103/PhysRevA.97.012333}
}

@Article{202200348,
  author   = {Zhou, Xiao-Dong and Wang, Sheng and Zhang, Hao and Zhang, Tian-Bing and Chen, Ye-Hong and Qin, Wei and Ning, Yu and Xia, Yan},
  journal  = {Ann. Phys.},
  title    = {Fast Generation of {2N}-Photon Fock States using Shortcuts to Adiabaticity and Ultrastrong Light–Matter Coupling},
  year     = {2023},
  number   = {1},
  pages    = {2200348},
  volume   = {535},
  doi      = {https://doi.org/10.1002/andp.202200348},
  url      = {https://onlinelibrary.wiley.com/doi/abs/10.1002/andp.202200348},
}

@article{Yu25,
author = {Jia-Wen Yu and Ke-Xiong Yan and Yuan Qiu and Jie Song and Ye-Hong Chen and Yan Xia},
journal = {Opt. Express},
number = {12},
pages = {26356--26372},
publisher = {Optica Publishing Group},
title = {Efficient generation of arbitrary photon-number-squeezed light via shortcuts to adiabaticity},
volume = {33},
month = {Jun},
year = {2025},
url = {https://opg.optica.org/oe/abstract.cfm?URI=oe-33-12-26356},
doi = {10.1364/OE.564732},
}

@article{wfj8-tgjz,
  title = {Efficient and flexible preparation of photonic NOON states in a superconducting system},
  author = {Li, Dong-Sheng and Kang, Yi-Hao and Shi, Zhi-Cheng and Xiao, Yang and Chen, Ye-Hong and Xia, Yan},
  journal = {Phys. Rev. A},
  volume = {113},
  issue = {1},
  pages = {012619},
  numpages = {12},
  year = {2026},
  month = {Jan},
  publisher = {American Physical Society},
  doi = {10.1103/wfj8-tgjz},
  url = {https://link.aps.org/doi/10.1103/wfj8-tgjz}
}

@article{PhysRevA.101.012345,
  title = {Deterministic interconversions between the Greenberger-Horne-Zeilinger states and the $W$ states by invariant-based pulse design},
  author = {Zheng, Ri-Hua and Kang, Yi-Hao and Ran, Du and Shi, Zhi-Cheng and Xia, Yan},
  journal = {Phys. Rev. A},
  volume = {101},
  issue = {1},
  pages = {012345},
  numpages = {11},
  year = {2020},
  month = {Jan},
  publisher = {American Physical Society},
  doi = {10.1103/PhysRevA.101.012345},
  url = {https://link.aps.org/doi/10.1103/PhysRevA.101.012345}
}
\end{document}